\RequirePackage{fix-cm}

\documentclass{svjour3}                                    % onecolumn (standard format)

\smartqed  % flush right qed marks, e.g. at end of proof

\usepackage{graphicx}
\usepackage[colorlinks,bookmarksopen,bookmarksnumbered,citecolor=red,urlcolor=red]{hyperref}
\usepackage{mathptmx}
\usepackage{amsfonts}
\usepackage{amsmath}
\usepackage{amssymb}
\usepackage{amsbsy}
\usepackage{mathptmx}      % use Times fonts if available on your TeX system
\usepackage{calrsfs}

\graphicspath{{PaperFigures/}}
\makeatletter
\newcommand{\BE}{\begin{equation}}
\newcommand{\EE}{\end{equation}}
\newcommand{\cond}{\big{|}}
\DeclareRobustCommand{\pder}[1]{%
\@ifnextchar\bgroup{\@pder{#1}}{\@pder{}{#1}}}
\newcommand{\@pder}[2]{\frac{\partial#1}{\partial#2}}
\DeclareMathAlphabet{\pazocal}{OMS}{zplm}{m}{n}

\usepackage{setspace}
\begin{document}
{\large 

\title{Sensitivity analysis and uncertainty quantification of 1D models of the pulmonary circulation 
%\thanks{Grants or other notes
%about the article that should go on the front page should be
%placed here. General acknowledgments should be placed at the end of the article.}
}
%\subtitle{Do you have a subtitle?     % if so, write it here}
\titlerunning{Sensitivity analysis and UQ in 1D models}         % if too long for running head

\author{Mitchel J. Colebank         \and
		M. Umar Qureshi         \and
		Mette S. Olufsen 
}
\authorrunning{Colebank, Qureshi, Olufsen} % if too long for running head

\institute{Mitchel J. Colebank \at
              North Carolina State University \\
              \email{mjcoleba@ncsu.edu} \\
           \and
           M. Umar Qureshi \at
           North Carolina State University \\
           \email{muquresh@ncsu.edu} \\
           \and
           Mette S. Olufsen \at
           North Carolina State University \\
           \email{msolufse@ncsu.edu}}

\date{Received: date / Accepted: date}
% The correct dates will be entered by the editor

\maketitle

\begin{abstract}
This study combines a one-dimensional (1D) model with micro-CT imaging and hemodynamic data to quantify uncertainty of flow and pressure predictions in the pulmonary arteries in a control and hypoxia induced hypertensive mouse. We use local and global sensitivity and correlation analysis to determine parameters that can be inferred from the model and available data. Least squares optimization is used to estimate mouse specific parameters, and Bayesian as well as asymptotic uncertainty quantification techniques are employed to determine confidence, credible, and prediction intervals for the model parameters and response. These techniques are used to examine the effects of network size and to understand how parameters change with disease (hypertension).  Results showed that the peripheral vascular resistance is the most sensitive, and as the network size increases the parameter behavior changes. Correlation analysis revealed that in hypertension large vessel stiffness is correlated with proximal resistance in the boundary. We were able to estimate identifiable parameters using both deterministic and Bayesian techniques (the maxima of the parameter distributions determined using Bayesian analysis aligned with local optima). From these estimates we determined confidence and prediction intervals, which all were within physiological expectation. Analysis of estimated parameter values for the representative mice studied here showed that the hypertensive mouse has stiffer (but larger) vessels and that compliance is decreased both in the proximal and peripheral vasculature.

\end{abstract}

\section{Introduction}
\label{intro}

One-dimensional cardiovascular models~\cite{Quateroni16,Shi11,Vosse2011} have been used to predict dynamics in both systemic \cite{Olufsen2000,Reymond2009}, pulmonary \cite{Qureshi2014}, and whole body networks \cite{Mynard15}.  This model type includes  three components: geometric specification of the vascular network, 1D approximation of the Navier-Stokes equations, and a constitutive equation relating pressure and cross-sectional area. Numerous studies have examined the effect on network size (the number vessels and junctions in the network) \cite{Epstein2015,Safaei2016}, boundary conditions (how to model the part of the network not represented explicitly)~\cite{Alalstruey08,Grinberg08,Olufsen2000,Qureshi2014}, and the constitutive equation (how to model wall distensibility)~\cite{Raghu11,Steele11,Valdez11}. Most of these have focussed on devising a framework to fit computed flow \cite{Olufsen2000,Reymond2009} and pressure \cite{Battista16} to measurements. More recently, these models have been expanded to determine uncertainty of model predictions~\cite{Arnold17,Eck16,Mirams16}. However, to our knowledge only a few studies~\cite{Brault16,Eck16,Gul16,vandeVosse11} have systematically examined the sensitivity of the model output to its parameters, essential to understanding how the system is modulated by disease.

Sensitivity analysis is commonly conducted in ODE models, computing sen\-si\-ti\-vi\-ties/im\-por\-tance of the model parameters to the model output (a function of the model states)~\cite{Campolongo2007,Ellwein2008,Marquis18,Pope2009,Wentworth2016}. This is more challenging in PDE models since A) they are more difficult to solve and B) the sensitivities are functions of both time and space. For the 1D fluid dynamics models studied here, sensitivities are computed at locations where data are available (in the main pulmonary artery) reducing the computational efforts needed to predict sensitivities.

In this study, we demonstrate how to compute sensitivities using both local  \cite{Ellwein2008,Marquis18} and global methods~\cite{Alexanderian2017,Gul16,Wu18} for a control and a hypoxia induced hypertensive mouse (hereafter referred to as hypoxic). The advantage of the former is that they are easy to compute, yet results are only valid in a region close to the parameter values at which they are evaluated. In comparison, global sensitivity methods determine how the model output varies over a specified parameter range and if parameters interact. However, they are more difficult to compute, and for non-linear models it is likely that given combinations of parameters cause the model to fail as they represent an unphysical domain. 

We use correlation analysis to determine an identifiable parameter set (a subset) that can be estimated uniquely given a model and available data~\cite{Miao11,Olufsen13}. Most complex non-linear models have a large number of parameters and limited data. As a result, it is likely that some parameters cannot inform the model predictions (insensitive parameters), while others may be correlated. Similar to sensitivity analysis, subset selection can be performed using both local and global methods. 

Subsequently, we  use Sequential Quadratic Programming~\cite{Paun2018} to minimize the least squares error between measured and predicted pulmonary arterial pressure. To examine these results we compute uncertainty of model predictions using both a local asymptotic and a global sampling type method \cite{Brady18,Marquis18} . These techniques will be used to analyze networks with 1, 3, and 21 vessels.  

Optimization results show that the model is able to fit data for both the control and hypertensive mice, independent of the network size.  As expected,  the hypertensive mouse has increased resistance and decreased compliance compared to the control mouse. Sensitivity analysis reveals that the peripheral vascular resistance is the most sensitive parameter,  which decrease with network size shifting the importance of determining total resistance from the periphery to the proximal vessels. The remaining parameters are less sensitive. An interesting observation is that as the networks grow the sensitivity to wall stiffness increases, while the sensitivity to peripheral compliance decreases,  in other words as the network grow it becomes easier to estimate wall stiffness compared to the peripheral compliance.

\section{Methods}\label{sec:Methods}

\subsection{Data}\label{Subsec:Data}
This study uses existing hemodynamic and micro computed tomography (micro-CT) imaging data from a control and a hypoxia induced hypertensive mouse. Detailed experimental protocols describing this data can be found in \cite{Tabima2012,Vanderpool2011}. All experimental procedures are approved by the University of Wisconsin Institutional Animal Care and Use Committee. The hemodynamics data include cycle averaged main pulmonary artery pressure and flow waveforms  gated to the ECG fiducial point  (see Fig.~\ref{fig:network}) \cite{Tabima2012} . We examine data from one control and one hypoxic male C57BL6/J  mouse  (12-13 weeks, weight~24 \,g) selected from groups of 7 control and 5 hypoxic mice. From each group we selected the mouse with flow and pressure waveforms closets to the group average.

\begin{figure}[h!]
\centering
\includegraphics[width=0.75\textwidth, angle=270]{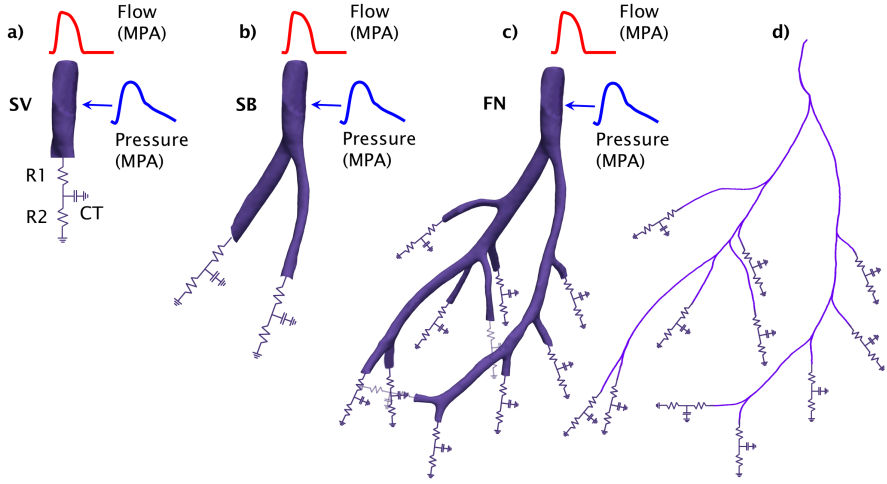}
\vspace{-8mm}
\caption{Mouse pulmonary networks constructed from micro-CT images from a control mouse. The three models considered are {\bf (a)} a single vessel model (SV), {\bf (b)} a single bifurcation model (SB), and {\bf (c)} a 21 vessel  model (FN). The 1D model {\bf (d)} was constructed by extracting centerlines from the 3D segmented network. The network is represented by a connectivity matrix with nodes and edges along with information on vessel radius and length.  At the network inlet an inflow waveform is given and a three element Windkessel model, relating flow and pressure is assigned for each terminal vessel.} \label{fig:network}
\end{figure}

The imaging data include stacked planar X-ray micro-CT images of pulmonary arterial trees from two male C57BL6/J mice, selected from groups of 8 control and 5 hypertensive mice (10-12 weeks, weight~24 \,g). The pulmonary arterial trees are imaged under a static filling pressure of 6.3 mmHg, while rotating the lungs in an X-ray beam at 1$^{\circ}$ increments to obtain 360 planar images. The Feldkamp cone-beam algorithm \cite{Feldkamp1984} was used to render the isometric 3D volumetric dataset (497$\times$497$\times$497 pixels) by reconstructing and converting the 360 planer images into Dicom 3.0. See Vanderpool et al. \cite{Vanderpool2011} for more details on animal preparation, handling  and experimental setup, and Karau et. al. \cite{Karau2001} for details on the micro-CT image acquisition.

\subsubsection{Network Geometry}

We developed a segmentation protocol, inspired by Ellewin et. al. \cite{Ellewin2015}, to extract vascular dimensions and network connectivity from the Dicom 3.0 files. This protocol uses  ITK-SNAP \cite{ITKSNAP} and Paraview (Kitware; Clifton Park, NY) to extract full 3D structure and the Vascular Modeling ToolKit (VMTK) \cite{VMTK} to obtain centerline coordinates and vessel  radii. 

As described in detail by Qureshi et al. \cite{Qureshi2018}, for each vessel, the unstressed radius ${r}_0$  is computed as a mean over slices $r_i$ away from the junction, and the vessel length $L$ is calculated as the sum of the shortest distances $l_i$ between successive points on the vessel, i.e.
\begin{equation}\label{eq:dist}
r_0 = \frac{1}{N_v}\sum_{i} r_i, \quad L = \sum_i l_i , \quad{\textrm{where}}\quad l_i = \|{\mathbf{ x}}_{i+1} - {\bf{ x}}_{i}\|, i = 0,\dots,N_v-1\\
\end{equation}
where $N_v$ is the total number of points along the centerline for vessel $v$. The coordinates of the shared junction between vessels is used to generate a connectivity map of the 3D structure.

For each mouse (control and hypoxic), we analyze three models: I a single vessel (SV - zero bifurcation) model, II a three vessel (SB - single bifurcation)  model, and III a 21 vessel (FN - full network- ten bifurcations) model. For both mice the models use the same connectivity map but the individual vessel radius and length vary as shown in Table~\ref{Tab:Network}. Figure \ref{fig:network} shows the three models for the control mouse. We included 21 vessels in the full network model as it was the most expansive network that can be identified with a one-to-one vessel map for both the control and hypoxic mice.  
\begin{table}
\centering
\caption{Vessel dimensions and connectivities for SV, SB, and FN networks for the control and hypoxic mice.}\label{Tab:Network} 
 {\footnotesize 
 \setlength\tabcolsep{3.5pt}
 \begin{tabular}{cccccc}
   \hline\noalign{\smallskip}
  \multicolumn{4}{c}{\bf \hspace{3cm}{Control}} & \multicolumn{2}{c}{\bf Hypoxia}\\ [2pt]
  \hline\noalign{\smallskip}
   Vessel & Connectivity & $r_0\times 10^{-1} $ &$ L\times 10^{-1}$ & $r_0\times 10^{-1} $ & $ L\times 10^{-1}$ \\
   Index   & (Daughters)   & (cm)               & (cm)           &  (cm)              & (cm)\\ 
   \noalign{\smallskip}\hline\noalign{\smallskip}
      1$^{\ast\dagger}$ & (2,3)	&  0.47  &  4.10 & 0.51& 3.58\\[2pt]
      2$^{\dagger}$ & (4,5)	&  0.26  &  4.45 &0.26& 4.03\\[2pt]
      3$^{\dagger}$ & (6,7)	& 0.37  &  3.72 & 0.37& 3.08\\[2pt]
      4 & (8,9)	&  0.24 &  2.41 & 0.25&2.92\\[2pt]
      5 &  --       &  0.13 &  0.52& 0.17 &0.65\\[2pt]
      6 & (14,15) &  0.32  &  2.02 & 0.28 &1.60\\[2pt]
      7 & --	       &  0.17  &  2.12 & 0.19 &0.93\\[2pt]
      8 & (10,11)&  0.23  &  3.11 & 0.24 &2.06\\[2pt]
      9 & --	       &  0.17  &  1.77 & 0.17 &0.51\\[2pt]
     10 &(12,13)&  0.20  &  2.62 & 0.22 &2.37\\[2pt]
     11 & --	      &  0.16  &  0.69 & 0.17 &0.88\\[2pt]
     12 & --	     &  0.15  &  1.40 & 0.19 &1.27\\[2pt]
     13 & --	     &  0.14  &  0.62 & 0.15 &0.51\\[2pt]
     14 &(16,17)  &  0.26  &  0.81 & 0.27 &1.20\\[2pt]
     15 & --	  &  0.19  &  1.84 & 0.19 &1.55\\[2pt]
     16 & (18,19)&  0.25  &  0.83 & 0.26 &0.71\\[2pt]
     17 &-- 	 &  0.15  &  3.02 & 0.18 &1.68\\[2pt]
     18 & (20,21)&  0.24  &  4.69 & 0.24 &3.55\\[2pt] 
     19 &-- 	&  0.15  &  1.77 & 0.18 &1.86\\[2pt]
     20 &-- 	&  0.22  &  1.78 & 0.23 &2.24\\[2pt]
     21 &-- 	&  0.18  &  0.55 & 0.19 &1.07\\[2pt] 
    \noalign{\smallskip}\hline\\[-6pt]
  \end{tabular}}\\
   {\footnotesize $\ast$ Dimensions for the single vessel (SV) model,  $\dagger$ the single bifurcation (SB) model, and the 21 vessel full network model  (10  bifurcations) (FN) model. For each connectivity pair $(i,j)$, $i$ refer to the left and $j$ to the right daughter.  Vessels marked with -- are terminal.}
\end{table}

\subsection{Fluid Dynamics Model}

Similar to previous studies of the pulmonary circulation~\cite{Olufsen2012,Qureshi2014}, the 1D fluid dynamics model is derived from the Navier-Stokes equations combined with a constitutive equation relating pressure and vessel area. The model predicts the pulmonary arterial pressure and flow by enforcing conservation of volumetric flow and axial momentum. Geometric and material properties as well as the in- and out-flow boundary conditions are extracted from imaging and hemodynamic data.

\subsubsection{The 1D Navier-Stokes equations}

Assuming that the vessels are cylindrical, that blood is incompressible, flow is Newtonian, laminar, and axisymmetric (with no swirl), and that the arterial walls are impermeable, conservation of mass and momentum \cite{Olufsen2000} gives
\begin{equation}\label{eq:mom}
\frac{\partial A}{\partial t} + \frac{\partial q}{\partial x} = 0,\quad \frac{\partial q}{\partial t} + \frac{\partial}{\partial x}\left(\frac{q^2}{A}\right)+\frac{A}{\rho}\frac{\partial p}{\partial x} = -\frac{2\pi\nu r}{\delta}\frac{q}{A},
\end{equation}
where $x$ and $t$ are the axial and temporal coordinates, $p(x,t)$ (mmHg) denotes the transmural blood pressure, $q(x,t)$ (ml/s) the volumetric flow rate, $A(x,t)=\pi r(x,t)^2$\,(cm$^2$) the cross-sectional area, and $r(x,t)$ (cm) the vessel radius. The blood density $\rho$\,(g/ml) and the kinematic viscosity $\nu$\,(cm$^2$/s) are assumed constant. The momentum equation is derived under the no-slip condition, satisfied by imposing a flat velocity profile over the lumen area~\cite{Olufsen2000} with a thin boundary layer that decreases linearly in the vicinity of the walls, where the transition to no-slip takes place.

\subsubsection{The constitutive equation}

The system of equations is closed by a constitutive equation (a wall model) relating pressure and cross-sectional area. This study uses a linear elastic wall model derived from balancing circumferential stress and strain~\cite{Safaei2016}. The wall model is derived under the assumptions that the vessels are cylindrical and that the walls are thin ($h/r_0\ll1$), incompressible, homogeneous, and purely elastic. We assume that the loading and deformation is  axisymmetric and that the vessels are tethered in the longitudinal direction. Under these conditions, the external forces reduce to stresses in the circumferential direction, yielding a linear stress-strain relation 
\begin{equation}\label{eq:linWall}
 p = \beta\left(\sqrt{\frac{A}{A_0}}-1\right),\quad\textrm{where}\quad \beta = \frac{Eh}{(1-\kappa^2)r_0}
\end{equation}
denotes the vessel stiffness. $E$ is  the circumferential Young's modulus, $\kappa=0.5$ is the Poisson ratio, $h$ is the wall thickness, and $A_0=\pi r_0^2$ (cm$^2$) refer to the undeformed cross sectional area \cite{Olufsen2000,Safaei2016} at $p=p_0=0$.

\subsubsection{Boundary conditions}\label{sec:BCs}

The 1D Navier-Stokes equations (\ref{eq:mom}) are hyperbolic with opposite characteristics, i.e. to be well-posed each vessel needs an inlet and an outlet boundary condition (see Fig.~\ref{fig:network}). We enforce this by specifying the flow at the network inlet (Fig.~\ref{fig:network}), conservation of flow and continuity of pressure at each junction, 
\begin{equation}\label{eq:bifcond}
p_p(L,t) = p_{d_i}(0,t) \quad\textrm{and}\quad q_p(L,t) = \sum_iq_i(0,t),
\end{equation} 
where the subscripts $p$ and $d_i$ ($ i = 1, 2$) refer to the parent and daughter vessels, and by attaching a three-element Windkessel model relating flow and pressure at each terminal vessel. The Windkessel model can be represented by a $RCR$ circuit relating pressure and flow as
\begin{equation}\label{eq:WK}
\frac{d p}{d t} -  R_1 \frac{d q}{d t} = q\left(\frac{R_1 + R_2}{R_2 C_p}\right) - \frac{p}{R_2 C_p},
\end{equation}
where $R_T= R_1 + R_2$ (mmHg\,s/ml)  is the total resistance, $R_1$ is the proximal,  $R_2$ the distal resistance, and $C_p$ is the total peripheral compliance of the vascular region in question  (see Fig.~\ref{fig:network}).

This system of equations is solved numerically, using a C++ implementation of the Lax-Wendroff two step method described in detail in \cite{Olufsen2000}.

\subsubsection{Nominal parameter values} \label{Subsec:Nominal}

The 1D model has three types of parameters specifying: network geometry, constants needed to specify the fluid and vasculature, in- and outflow boundary conditions. Some of these can be measured or found in literature, while others must be estimated. 

\paragraph{Network geometry} (vessel length, radius, and connectivity) is extracted from imaging data as described in Sec. \ref{Subsec:Data}.  While these quantities carry uncertainty, in this study, we assume that all geometric properties are constant and known.

\paragraph{Fluid  and vascular properties:} The blood density $\rho=1.057$\,g/ml \cite{Riches1973},  the kinematic viscosity $\nu= 0.0462$\,cm$^2$/s, measured at  a shear rate of 94\,s$^{-1}$ \cite{Windberger2003}, and the boundary layer thickness  $\delta = \sqrt{2\pi\nu/T}$ \cite{Vosse2011} are assumed constant.The wall stiffness  $\beta$, approximated as
\begin{equation}\label{eq:betaEst}
\beta = \frac{2(A_0 Z_c)^2}{\rho},
\end{equation}
where the characteristic impedance $Z_c$ is estimated from the slope of  the pressure-flow loop including 95\% of the flow during ejection phase \cite{Qureshi2017}, is allowed to vary. 

\paragraph{Inflow condition:} We specify flow into the network using the measured flow waveform along with the length of the cardiac cycle $T=$\,1/HR (s). Both are kept constant for all simulations.

\paragraph{Outflow condition:} A Windkessel model (\ref{eq:WK}) relating pressure $p$ and flow  $q$ as a function of vascular resistance and compliance is attached to each terminal vessel. This model is formulated using 3 parameters relating vessel compliance, and resistance $\theta_{wk}=\{R_1,R_2,C_p\}$. These parameters are functions of the total peripheral resistance $R_{T} = \overline{p}/\overline{q}$, the flow distribution in the network, and the global time constant $\tau =R_{T}C_{p}$. 

For the SV model $R_T = \overline{p}/\overline{q} = R_1 + R_2$. Similar to previous studies  \cite{Battista16,McDonald:UK}  we assume $R_{1}  = 0.2 R_{T}$, i.e. and $R_{2} = R_{T}-R_{1}$.  As suggested by Stergiopulos et al. \cite{Stergiopulos1995}, peripheral compliance $C_p = \tau/R_T$, where the time-constant $\tau$ is estimated by fitting the diastolic pressure decay $p_d(t)$ to an exponential function
\begin{equation}
p_d(t) = p(t_d)\exp(-(t-t_d)/\tau),
\end{equation}
where $t_d$ denotes the onset of diastole.
 
For models with more than one vessel (SB and FN), $\theta_{wk}$ are estimated by distributing $R_T$ to each of the terminal vessels $j$ as
\begin{equation}\label{WKnetwork}
  R_{Tj} = \frac{\overline{p}}{\overline{q}_j},\quad \textrm{and}\quad C_{pj}=\frac{\tau}{R_{Tj}},
\end{equation}
where $\overline{q}_j$ is the mean flow to vessel $j$, determined by applying Poiseuille's law recursively at each junction, giving
\begin{equation}\label{eq:WKRnom}
  \overline{q}_{d_i}  =  \frac{ G_{d_i}}{\sum_i G_{d_i}}\overline{q}_p,\ \textrm{where} \ G_{d_i} = \left(\frac{\pi r_0^4}{8\mu L}\right)_{d_i} \  \textrm{for} \quad  i = 1,2.
\end{equation}
Here $\overline{q}_{d_i}$ denotes the mean flow to daughter vessel $i$. Similar to the SV model, the total resistance is distributed as  $R_{1j}  = 0.2R_{Tj}$ and $R_{2j} = R_{Tj}-R_{1j}$.

In summary, each outlet $j$ requires specification of 3 parameters $(R_{1j}, R_{2j}, C_{pj})$, i.e. the SV model has 3 outflow parameters, the SB model has 6 outflow parameters, and the FN model has 33 outflow parameters.

\paragraph{Summary:} 
The model parameters can be grouped into two categories: parameters needed to set up the conservation equations $\theta_h = \{T, \nu, \rho, \delta, \beta_i \}$ for $i=1..N$, where $N$ is the number of vessels, and parameters needed to specify the outflow boundary condition $\theta_{wk} = \{R_{1j}, R_{2j}, C_{pj}\}$ for $j=1..M$, where $M$ is the number of terminal vessels, i.e. the SV, SB, and FN models have 8, 21, and 138 parameters, respectively. To reduce the number of parameters, we assume that the length of cardiac cycle, viscosity, density, and boundary layer thickness are constant, while parameters representing vessel stiffness $\beta$, and outflow boundary conditions $\theta_{wk} = \{R_1, R_2, C_T\}$ vary. This still leaves 54 parameters to estimate in the FN model. To reduce the parameter set, similar to \cite{Paun2018,Qureshi2018}, we assume that vessel stiffness $\beta$ is constant throughout the network \cite{Krenz2003} and introduce global scaling factors $r_1,r_2$ and $c_1$ for the Windkessel model, i.e. for any terminal vessel $j$
\[
\widehat{R}_{1j} = 
\widehat{r}_1R_{1j}, \quad \widehat{R}_{2j} = \widehat{r}_2R_{2j},\quad \widehat{C}_{pj} = \widehat{c}_1C_{pj},\]
where $\widehat{.}$ indicates the optimized quantities, i.e., the final parameter set analyzed is 
$\tilde{\theta} = \{\beta,r_1,r_2,c_1\}$.

\subsection{Parameter Estimation}

Estimated parameters $\hat{\theta}$ are predicted by minimizing the least squares error 
\begin{eqnarray} \label{eq:J}
 &&\min_{\hat{\theta}} J = \min_{\hat{\theta}}  \left(\frac{r^T r}{N-P}\right),   \\
 &&r =  \left(p_m(0,t_1) - p_{data}(t_1), p_m(0,t_2)-p_{data}(t_2), ..., p_m(0,t_N)-p_{data}(t_N)\right), \nonumber
 \end{eqnarray}
 where $N$ is the number of data points and $P$ is the number of estimated parameters. $p_{data}(t_i)$ is the main pulmonary artery pressure data measured at $t=t_i$, $p_m(0,t_i)$ denotes  the model prediction of the pressure at the root of the main pulmonary artery for a given value of $\theta$.

\section{Model Analysis}\label{sec:ModelAnalysis}

To determine how the identifiable parameters  $\hat{\theta}$  influence fluid dynamics predictions we conduct sensitivity and correlation analysis, parameter estimation, and uncertainty quantification.

\subsection{Sensitivity Analysis}

Local sensitivity analysis analysis~\cite{Ellwein2008} is done using a derivative based method, while global analysis is done using Morris screening~\cite{Sumner2012}.

\subsubsection{Local sensitivity analysis} \label{Subsec:localsens}

Derivative based sensitivity analysis is one of the most common methods for computing local sensitivities of the model output to its parameters in a small neighborhood of given parameter values \cite{Olsen2015,Pope2009}. 
For a given quantity of interest $y = f(t;\theta)$, the sensitivity $S_i$ of $y$ to the parameter $\theta_i$ at time $t$ is defined as 
\BE \label{eq:localsens}
S_i(t; \theta) = \pder{y}{\theta_i} = \pder{f(t;\theta)}{\theta_i},
\EE
where the index $i \in \{1,2,3,4\}$ refers to the parameter of interest.

In this study, the wall stiffness parameter $\beta$ is of order $10^4$ and $10^5$ for the control and hypoxic models, respectively, while the scaling parameters $r_1, r_2, c_1$ are of order $10^0$. To obtain parameters of the same order of magnitude, we apply log-based scaling introducing a scaled parameter vector $\tilde{\theta} = \log(\theta)$. 
The advantage of this scaling is that sensitivities become relative to the parameters, 
\BE
\tilde{S}_i(t; \theta) = \pder{y(t;\theta)}{\tilde{\theta_i}}= \frac{\partial y(t;\theta)}{\partial \theta_i} \theta_i, \ \ \ i=1... N. 
\label{eq:relsensres}
\EE
There are several ways to compute the sensitivity matrix  (\ref{eq:relsensres}). If the model output $y$ is an analytical function of $\theta$, sensitivities can be determined directly by differentiation. If not, and the number of parameters is small, explicit sensitivity equations can be derived~\cite{Ellwein2008}.  More commonly, sensitivities are approximated  using finite differences (FD) (e.g. using backward Euler or a centered difference scheme), Automatic Differentiation~\cite{Ellwein2008,Griewank1989}, or complex step methods~\cite{Banks2015,RalphBook}. The advantage of FD methods is that they are easy to compute, yet the step size for the parameter perturbation is limited by the accuracy of the numerical solver \cite{Lott1985}. The other methods are more accurate, but require tedious computations. In this study, we compute local sensitivities using the  centered difference FD scheme
\BE 
\pder{y}{\theta_i} \approx \frac{f(t;\theta + \bold{e}_i h )- f(t;\theta - \bold{e}_i h)}{2h},
\label{eq:FD}
\EE
where $h$ is the step size and $\bold{e_i}$ is a unit vector with $0$'s everywhere except the $i$-th entry. 

The length of the cardiac cycle $T = 0.11s$ and the number of time steps per period is 8192, giving $h= \sqrt{\Delta t}  \approx 10^{-3}$ for the second-order accurate Lax-Wendroff scheme.

Local sensitivities are functions of time. Given the periodic model output, it is suitable to compute ranked sensitivities $\overline{S_i}$ by averaging over time using the 2-norm
\BE
\overline{S_i} = || \tilde{S}_i ||_2.
\EE

\subsubsection{Global sensitivity analysis} \label{Subsec:morris}

Global sensitivity methods can be categorized in two classes: variance based methods and screening. Variance based methods, including Sobol' indices, quantify parameter importance and interaction by imposing a break down of the total variance in an Analysis of Variance manner \cite{Wentworth2016}. However, Sobol' indices are computationally expensive, they require $N\cdot(P+2)$ model evaluations, where $P$ is the number of parameters in the model and $N$ is the number of samples \cite{Saltelli2008}. Screening methods, such as Morris' screening, require substantially less function evaluations, but compute sensitivities on a coarser scale. While Morris' screening provides less information about parameter sensitivities compared to Sobol methods, Campolongo et. al. \cite{Campolongo2007} have shown that screening based methods agree well with the total sensitivity measure $S_T$ obtained from Sobol' indices. Thus the Morris' indices are a valid tool for measuring global parameter sensitivity. 

Given the complexity of the fluid dynamics model, we pursue Morris' screening methods \cite{Morris1991}, which involves the computation of so called ``elementary effects", determining the relative change in model output to a relative change in parameter values. The main difference between local and screening methods is that the latter samples parameters throughout the parameter space computing the local sensitivities at these points. The screening method quantifies the effects of a parameter $\theta_i$ on the output quantity as  a) negligible overall, b) linear and additive, or c) having nonlinear effects or involved in higher order interaction with other parameters.

To perform global sensitivity analysis, parameters are mapped from their bounded parameter space $\Theta$ to the unit hypercube $[0,1]^P$ where $P$ is the number of parameters of interest. For the fluid dynamics model, we assume that parameter bounds are known but that little information about the shape of the parameter distribution is available.  For such systems it is reasonable to assume that the \textit{a priori} parameter distribution is uniform within given upper and lower bounds.

The elementary effects are computed as
\BE \label{eq:EE1}
d_i(\theta) = \frac{f(\theta + \bold{e}_i \Delta) - f(\theta)}{\Delta},
\EE
where $\bold{e}_i$ is the unit vector in the $i$-th direction, $i = 1, 2, \dots, P$. The step size $\Delta$ is chosen from the set $\Delta \in \{1/(L-1), 2/(L-1) \dots, (L-2)/(L-1) \}$, where $L$ denotes the number of levels at which parameters are perturbed. To compute the elementary effects we sample $K$ parameter values from the uniform distribution for the parameter $\theta^j_i$, giving
\BE \label{eq:EE2}
d^j_i(\theta_i)  = \frac{f(\theta^j + \bold{e}_i \Delta, t) - f(\theta^j)}{\Delta} ,  \ \ j = 1,2,\dots, K.
\EE

As has been noted in multiple studies~\cite{Campolongo2007,Morris1991,RalphBook}, symmetry of parameter distributions can be preserved by choosing  $L$ as an even number. We use the algorithm by \cite{Wentworth2016} to compute each parameters elementary effect using the model output $y_m(t;\theta)$ and the step size $\Delta$ scaled by the magnitude of each parameter.

The average response and variance is obtained by integrating outcomes from multiple iterations. To obtain scalar valued quantities for the global analysis, we take the two norm of the absolute value of the elementary effects. The modified Morris' indices are
\BE 
\mu^*_i = \frac{1}{K} \sum_{j=1}^K |d_i^j|, \hspace{7mm} \mu_i = \frac{1}{K} \sum_{j=1}^K d_i^j,  \hspace{7mm} \sigma^2_i = \frac{1}{K-1} \sum^{K}_{j=1} \left( d_i^j - \mu \right)^2. 
\label{eq:morris}
\EE

Here $\mu^*$ quantifies the individual effect of the input on the output, i.e. the sensitivity of the model with respect to the parameter selection. The variance estimate describes the variability in the model response due to parameter interactions, e.g. parameters with a large $\mu^*$ and $\sigma^2$  have large effects on the model output and are highly nonlinear in the model. Similar inference can be made for other combinations of $\mu^*$ and $\sigma^2$. The use of $\mu^*$ rather than $\mu$ is discussed at length in \cite{Campolongo2007}, and has been shown to be a better indicator of sensitivities. However, since we take the two norm of our elementary effects, the values of $\mu*$ and $\mu$ are the same in this study. The quantities $\mu^*$ and $\sigma^2$ can be used in combination to determine which parameters are the most sensitive in the system and develop a parameter ranking.

For the randomized Morris' algorithm, the number of samples $r$ is set to 50, the number of levels of the parameter space $L = 20$, and the step size $\Delta = \frac{L}{2(L-1)} \approx 0.526$. 

\subsection{Correlation Analysis} \label{Sec:correlation}

To identify parameter correlations, we analyze the covariance matrix $C$, which for constant model variance $\sigma^2$, can be approximated asymptotically from the sensitivity matrix $\tilde{S}$~\cite{Banks2013} as
\BE \label{eq:covariance}
  C = \sigma^2\left(\tilde{S}(t;\theta)^T \tilde{S}(t;\theta)\right)^{-1}, \hspace{5mm} \sigma^2 = J,
\EE
where $J$ is the least squares cost (\ref{eq:J}). We calculate correlations as
\BE \label{eq:correlation}
  c_{ij} = \frac{C_{ij}}{\sqrt{C_{ii}\,C_{jj}}},
\EE
where $c_{ij}$ is an upper triangular symmetric matrix with diagonal elements $c_{ii} = 1$ 
and $|c_{ij}| \leq 1$. 
We use the structural correlation method~\cite{Olufsen13}, which remove least sensitive parameters for which $|c_{ij}| > \gamma = 0.9$. 

\subsection{Optimization}\label{Subsec:Optimization}

Identifiable model parameters are estimated using the the Sequential Quadratic programming (SQP) algorithm, minimizing (\ref{eq:J}) within specified parameter bounds. The function \texttt{fmincon.m} was used in MATLAB, with parameter bounds $\Theta_{control} = [10^4,10^5]\times [0.05,3]^9$ and $\Theta_{hypoxic} = [10^5,10^6]\times [0.05,3]^9$ for the control and hypoxic mice, respectively. These bounds were chosen within physiological limits to ensure that the model will successfully compute without crashing.
 
Optimization was run on an iMac (3.4 GHz Intel Core i7, 16GB RAM, OS 10.13.4). For each optimization, eight initial values are sampled from a uniform distribution with fixed upper and lower bounds. The algorithm is iterated until the convergence criterion was satisfied with a tolerance {$\epsilon <10^{-8}$}.

\subsection{Uncertainty Quantification}

Uncertainty quantification includes a broad class of techniques that analyze the predictive nature of a mathematical model. There are two main types of uncertainty: \textit{aleatoric} uncertainty (or statistical uncertainty), which refer to the inherent noise in the experimental observations, and \textit{epistemic} uncertainty (or scientific uncertainty), which refer to the uncertainty of the model, including modeling assumptions and/or lack of knowledge of the physical process \cite{RalphBook}.  In this study, we focus on \textit{epistemic} uncertainty by analyzing the effects of model parameters on the model predictions of pulmonary arterial pressure.

Both frequentist and Bayesian techniques can be used to determine intervals on which parameters are most likely to lie, and use these to determine the range of the expected model output. In this study, we use both the frequentist and Bayesian framework to construct intervals around the model prediction and around the parameter values and distributions.

\subsubsection{Frequentist analysis}

In the frequentist framework, confidence and prediction intervals are constructed for both the parameters and model prediction. To compute the confidence intervals for a given parameter, we consider the estimated parameters $\hat{\theta}$ minimizing the residual vector $r(t;\theta)$ defined in (\ref{eq:J}) and the associated local sensitivity matrix $\hat{S} = \tilde{S}(t;\hat{\theta})$.  Using this framework and the variance $\hat{\sigma}^2 = \hat{J}$ (\ref{eq:J}), the parameter confidence interval can be determined as
\BE \label{eq:freqCI}
\theta^{CI}_{i} \equiv \left[\hat{\theta_i} - t_{n-p}^{1-\alpha/2} \hat{\sigma} \sqrt{\left(\hat{S}^T \hat{S} \right)^{-1} }, \hat{\theta_i} + t_{n-p}^{1-\alpha/2} \hat{\sigma} \sqrt{ \left(\hat{S}^T\hat{S}\right)^{-1} } \ \ \right],
\EE
where $\alpha = 0.05$ for the $t$-statistic with $N-P$ degrees of freedom. 

The confidence interval for the model response ($G_i$) can be computed as
\BE \label{eq:freqCImodel}
y^{CI}(t_i) \equiv \left[y(t_i; \hat{\theta})  - t_{n-p}^{1-\alpha/2} \hat{\sigma}  \sqrt{\Gamma_i}, \ \ y(t_i; \hat{\theta})  + t_{n-p}^{1-\alpha/2} \hat{\sigma} \sqrt{\Gamma_i} \ \ \right],
\EE
where  $\Gamma_i = G_i^T\left(\hat{S}^T\hat{S}\right)^{-1}G_i$ and 
\[
G_i^T = \left(\pder{y(t_i; \hat{\theta})}{\theta_1}, \dots, \pder{y(t_i; \hat{\theta})}{\theta_k} \right).
\]

The prediction intervals for the model response are calculated as
\BE \label{eq:freqPImodel}
y^{PI}(t_i) \equiv \left[ y(t_i; \hat{\theta})  +t_{n-p}^{1-\alpha/2} \hat{\sigma} \sqrt{1 + \Gamma_i}, \ \ y(t_i; \hat{\theta})  - t_{n-p}^{1-\alpha/2} \hat{\sigma} \sqrt{1 + \Gamma_i} \ \ \right].
\EE

\subsubsection{Bayesian analysis}

In contrast to the frequentist perspective, Bayesian intervals (credible and prediction) are computed from the posterior distributions of the parameters in question. In this study, we employ the Delayed Rejection Adaptive Metropolis (DRAM) algorithm~\cite{Haario2006,Laine2007} to determine posterior parameter distributions, credible and prediction intervals. Results from DRAM simulations are also used to study if estimated parameters  are pairwise correlated. The latter can be  done via pairwise plots of the distributions. The advantage of DRAM is that if used with enough samples, clear global trends can be extracted. However, this method is computationally intensive making it difficult to carry out on multiple datasets. Hence we focus on comparing the sampling results with asymptotic estimates, which can be easily computed. 

In this framework, the posterior parameter densities are computed as
\BE \label{eq:bayes}
\pi\left(\theta \cond y \right) = \frac{\pi\left( y \cond \theta\right) \pi_0(\theta)}{\pi(y)} = \frac{\pi\left( y \cond \theta\right) \pi_0(\theta)}{\int_{\mathbb{R}^P}\pi\left(y\cond\theta\right) \pi_0(\theta)d\theta}
\EE
given the data $y$. We assume that the prior distribution $\pi_0(\theta)$ is  non-informative (i.e. flat), that the likelihood function $\pi\left( y \cond \theta\right)$ in (\ref{eq:bayes}) can be specified, and that it displays statistically properties of the data. We assume that the model error is independent and identically distributed (iid) with mean zero and constant variance $\sigma_\epsilon^2$, $\epsilon_i \sim \mathcal{{N}}(0,\sigma_\epsilon^2)$. Under these assumptions the likelihood function is
\BE\label{eq:likelihood}
\pi\left( y \cond \theta\right) = L\left( \theta,\sigma_\epsilon^2 \cond y\right) = \frac{1}{\left( 2\pi \sigma_\epsilon^2 \right)^{N/2}}e^{-SSE / 2\sigma_\epsilon^2}.
\EE
Subsequently, credible and prediction intervals for the model response are established by taking $M$ samples from the posterior distributions \cite{Haario2006,RalphBook}.

\section{Results} \label{sec:Results}

We use sensitivity and covariance analysis to determine a parameter subset that can minimize the least squares error between model predictions and data. Estimated parameters are interpreted to discuss effects  of network size and disease.

\subsection{Sensitivity Analysis}

Both local and global sensitivity analysis of the model output with respect to the nominal parameter values $\theta = \{\beta, r_1, r_2, c_1\}$ show that all parameters impact model predictions independent of the network size and/or disease.  Sensitivities, depicted  in Figs. \ref{fig:sensitivity}, \ref{fig:globalsensitivity}, and \ref{fig:rank}, are mapped onto $[0,1]$ by dividing by the maximum sensitivity for each network.  The analysis shows that for all networks and both mice $r_2$ is  the most sensitive parameter. Analysis of the  remaining parameters show that as more branches are added, vessel stiffness $\beta$ becomes more sensitive, while the proximal resistance $r_1$ and compliance $c_1$ scaling factors become less sensitive, indicating that the sensitivity to proximal resistances $R_1$ and peripheral compliances $C_p$ is reduced.
The sensitivity to $c_1$  separates the two groups. For the hypoxic mouse $c_1$ is more sensitive than $\beta$, which is opposite in the control group.

Analysis of the time-varying sensitivities show that $r_1$ and $\beta$ are more sensitive during systole, while $r_2$ is more sensitive during diastole. Sensitivities for $r_2$ and $c_1$ do not vary significantly over the cardiac cycle.  

Plots of the average elementary effects compared to the standard deviation are shown in Fig. \ref{fig:globalsensitivity}. Overall, results agree with the local analysis. In addition to  ranking, the global analysis provides a measure of nonlinearity and/or parameter interaction via $\sigma$. Results show that $r_2$ has the largest interaction effect, while $r_1$ has the smallest $\sigma$, regardless of the mouse or model. While nonlinear/interaction effects from $\beta$ increase with network size.

In summary, the global analysis shows that the stiffness parameter $\beta$ (for both control and hypoxic mice) becomes more sensitive as more vessels are added. Moreover, the compliance scaling factor $c_1$ has a high $\sigma$ across all three model types. These results show that the sensitivity of the compliance is highly dependent on magnitudes of the other parameters in the system.

\begin{figure}[ht]
\centering
\includegraphics[width=0.75\textwidth, angle=270]{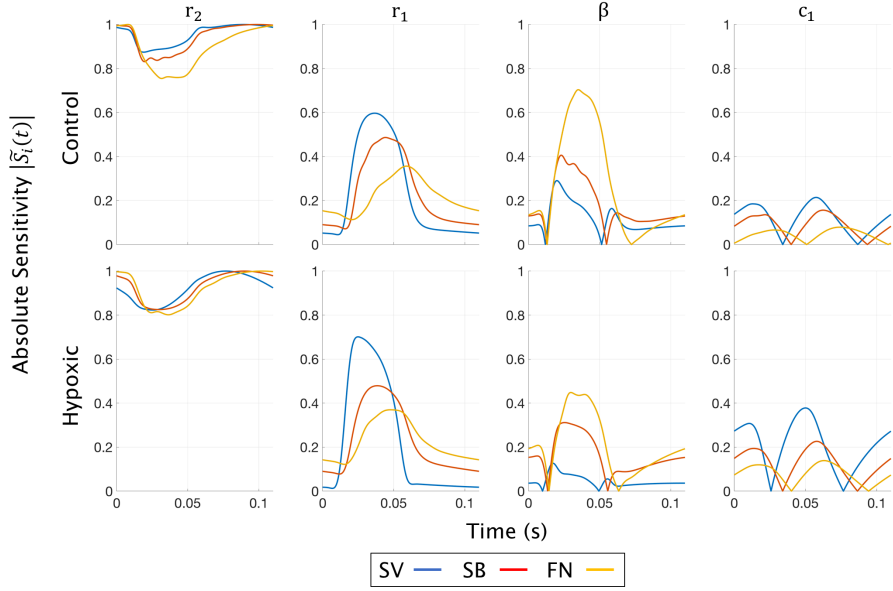}
\vspace{-8mm}
\caption{Normalized local sensitivities in the root of the main pulmonary artery pressure at each time point in the cardiac cycle. $r_2$ exhibited sensitivities that were largest in magnitude, while the other parameters were of smaller magnitude.}
\label{fig:sensitivity}
\end{figure}

\begin{figure}[ht]
\centering
\includegraphics[width=0.75\textwidth,angle=270]{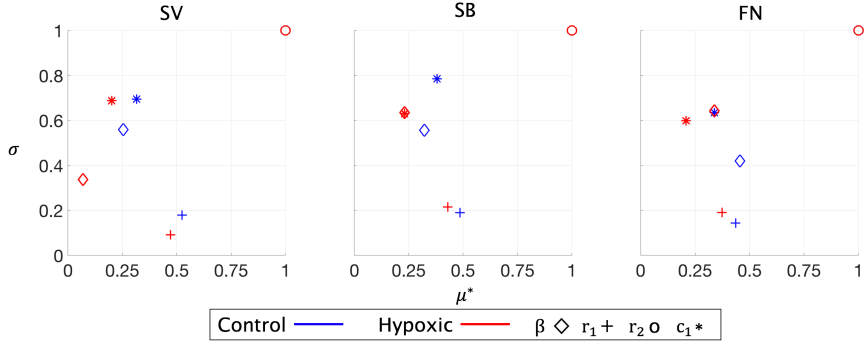}
\vspace{-20mm}
\caption{Global sensitivity results: normalized Morris' indices $\mu^*$ and $\sigma$ for the three models in the control and hypoxic mice. For both mice, $r_2$ has the largest value of $\mu^*$ and $\sigma$. The global sensitivities show that the parameter $\beta$ increases in sensitivity as more vessels are added to the network.}\label{fig:globalsensitivity}
\end{figure}

\begin{figure}[ht]
\centering
\includegraphics[width=0.75\textwidth,angle=270]{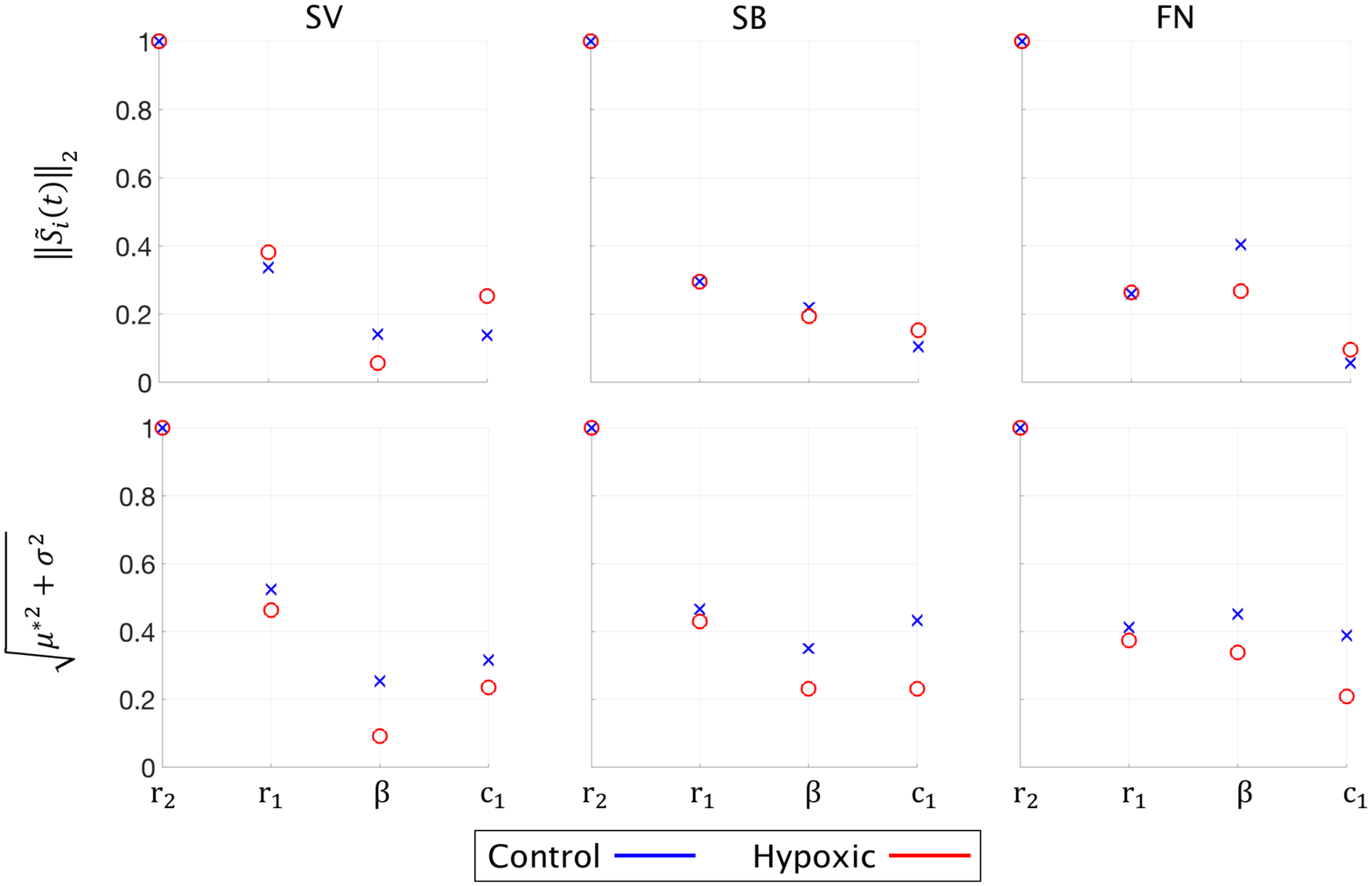}
\vspace{-8mm}
\caption{Ranking of parameters based on their local and global sensitivity results. All ranking metrics were scaled to the rank value of $r_2$, which was largest in magnitude.}\label{fig:rank}
\end{figure}

\subsection{Correlation Analysis}

The local parameter sensitivities were used to compute the covariance matrix, given in (\ref{eq:covariance}) for each model. The control models did not produce any correlations that were larger than the threshold $\gamma=0.90$. All three networks in the hypoxic model label $\beta$ and $r_1$ as correlated. To explore this correlation over a larger parameter space, we conducted a DRAM simulation on each sub-model. Results, shown in Fig. \ref{fig:pairwise}a for the hypoxic FN,  support the covariance based analysis that $\beta$ and $r_1$ are correlated. In fact, this figure suggests that all parameter pairs are correlated. Fixing $\beta$ at its nominal value removes all correlations as shown on Fig. \ref{fig:pairwise}b. While correlations were the strongest for the hypoxic FN model, the effects of correlation persisted for all networks.

\begin{figure}[ht]
\hspace{-0.4cm}
\includegraphics[scale=0.2]{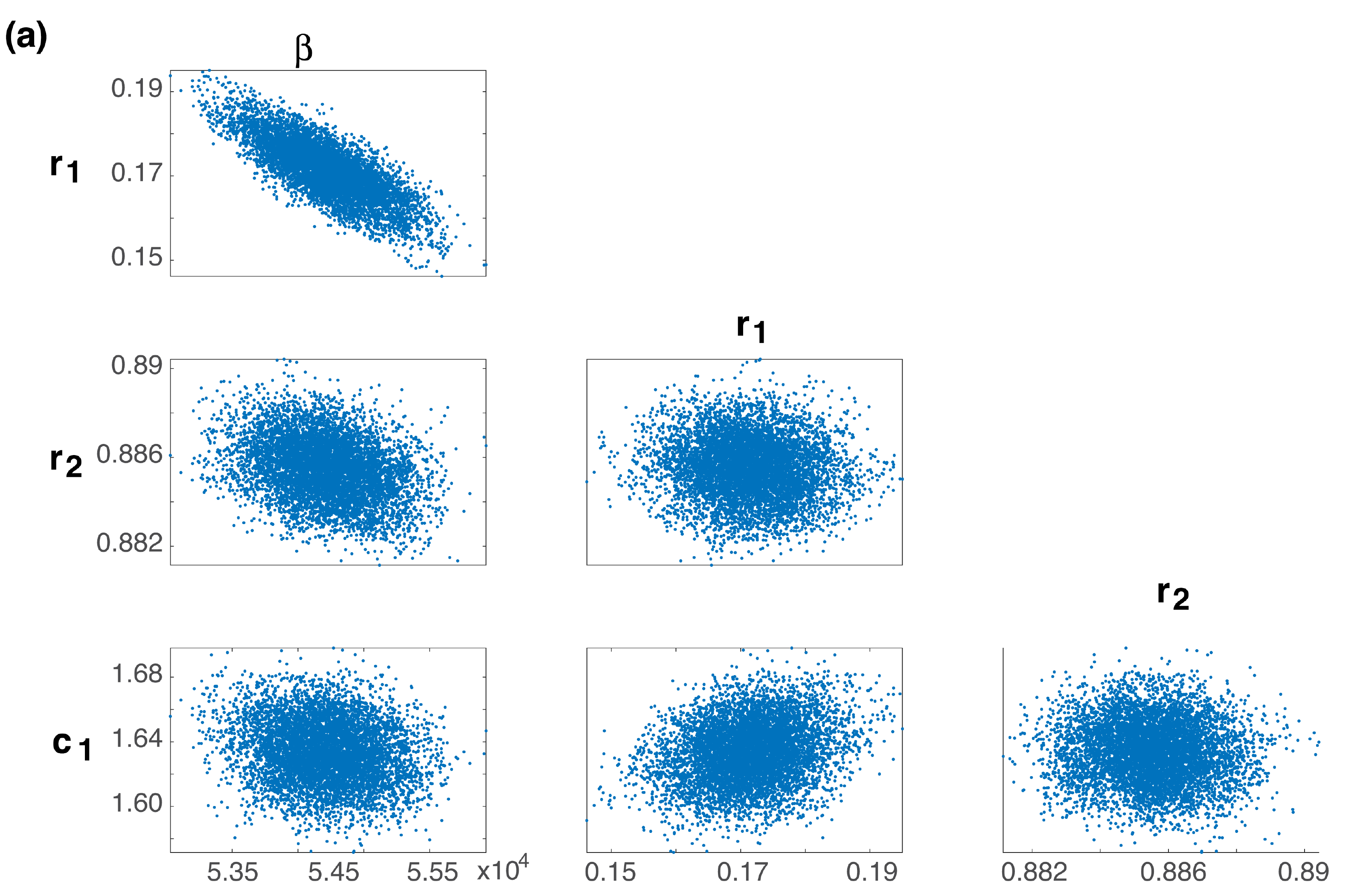}\hspace{-0.5cm}
\includegraphics[scale=0.2]{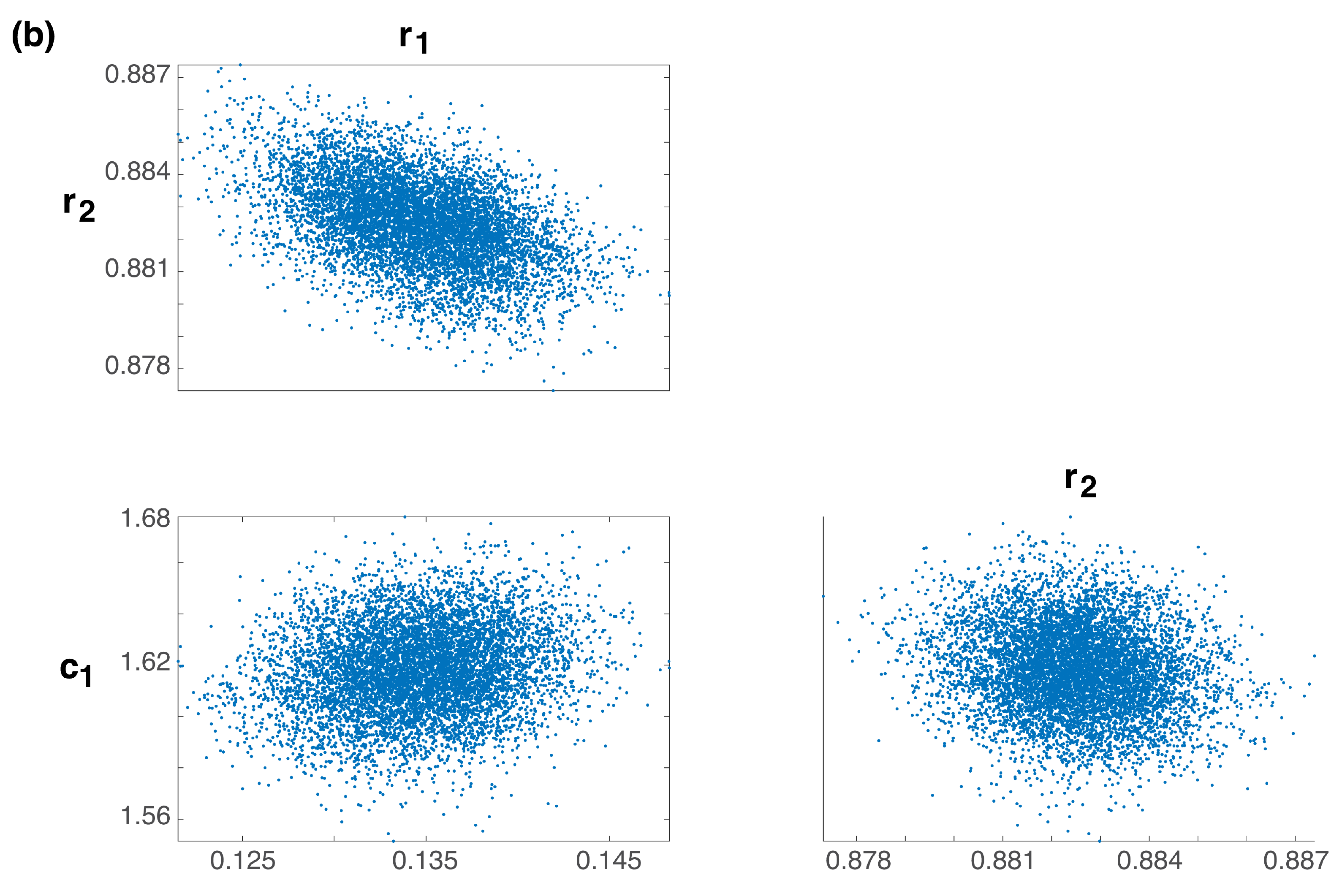}
caption{{\bf(a)} Pairwise plots from 10,000 iterations in DRAM for the FN model; {\bf(b)} Pairwise plots from DRAM when $\beta$ was fixed at its nominal value. {\bf(a)} shows that the parameters $\beta$ and $r_1$ are highly correlated, as was indicated by the structured correlation results. By fixing $\beta$ (panel {\bf(b)}), the parameters become less correlated.}\label{fig:pairwise}
\end{figure}

\subsection{Optimization Results}

Results of subset selection showed that for some datasets, $\beta$ is correlated with $r_1$. However, the correlation was not strong for all sets. Assuming that wall stiffness does not change as branches are added, we estimated one value of $\beta$ over all three networks, i.e. we estimated a total of 10 parameters
\BE
\theta_{0} = \{\beta, r^{SV}_1,r^{SV}_2,c^{SV}_1, r^{SB}_1,r^{SB}_2,c^{SB}_1, r^{FN}_1,r^{FN}_2,c^{FN}_1 \}  = \{\beta, \theta^{SV}, \theta^{SB}, \theta^{FN}\}
\EE
minimizing
\BE\label{eq:initialcost}
J^* = \frac{1}{3N-10}{r^*}^T {r^*}, \ \ \ \ \text{where} \ \ \ \ 
r^* = \begin{bmatrix} \bold{y}^{SV}(t; \theta^{SV}) - \bold{y}_{data}(t)\\\bold{y}^{SB}(t; \theta^{SB}) - \bold{y}_{data}(t)\\ \bold{y}^{FN}(t; \theta^{FN}) - \bold{y}_{data}(t) \end{bmatrix},
\EE
where $\bold{y}_{data}(t)$ denotes the output data.

Eight initial values are sampled from a uniform distribution on the parameter space given in Sec. \ref{Subsec:Optimization} to ensure convergence. Optimal model solutions are depicted in Fig. \ref{fig:PressurePlots} and the optimal parameter values are given in Table~\ref{Tab:Optimized}. Subsequently, DRAM simulations were conducted with $\beta$ fixed using the optimized values to determine {\it a priori} parameter distributions. 

Results depicted in Fig.~\ref{fig:PressurePlots} show that all models fit the data well. Overall, the hypoxic model gives a lower cost than the control model, which is on the order of $10^{-2}$ vs.$10^{-1}$, respectively. The cost across different models does not change in order of magnitude, indicating that similar predictions are obtainable with different sized networks. One key difference between the control and hypoxic predictions is that the hypoxic model accurately predicts the systolic rise and diastolic decay of the pressure curve. In contrast, the control predictions are unable to capture the exact shape of the pressure wave.
  
Table~\ref{Tab:Optimized} shows results for the optimized values of the ratio $R_1 / R_T$, the total peripheral resistance and compliance  $R_T$ and $C_P$. The $R_1/R_T$  ratio decreases as the number of vessels in the network is increased for both the control and hypoxic mouse. The total resistance $R_T$ was consistently higher in the hypoxic mouse,  whereas the compliance $C_P$ was higher for the control mouse. For both the control and hypoxic mice the total resistance $R_T$ decrease as more vessels were added to the network. 

\begin{figure}[ht]
\centering
\includegraphics[width=0.75\textwidth,angle=270]{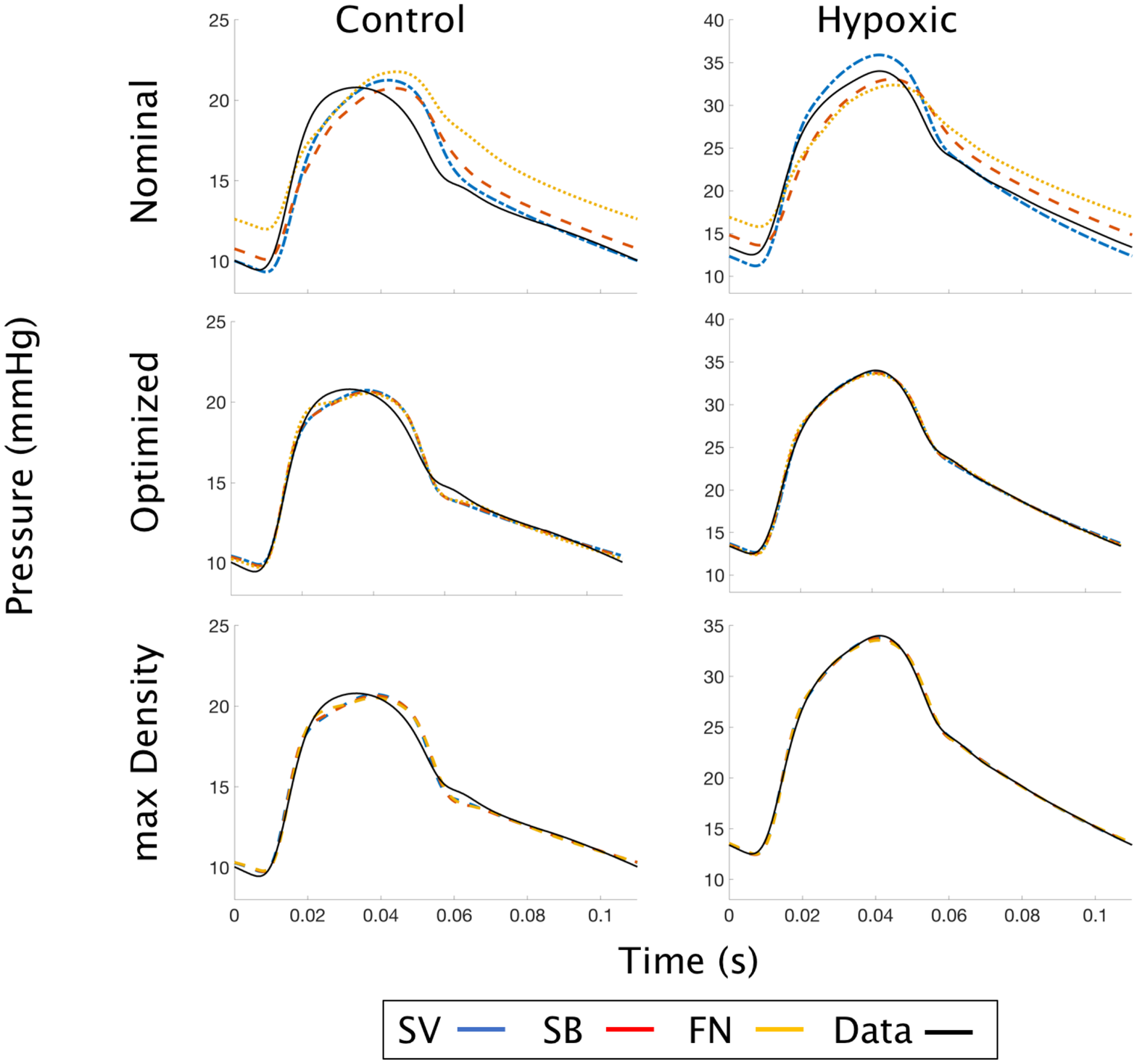}
\caption{The main pulmonary artery pressure data against model predictions using nominal parameter estimates. The control (left column) and hypoxic (right column) models are plotted against the data using the nominal parameters (top row), optimized parameter (middle row), and maximum density parameters obtained from posterior densities (bottom row).}\label{fig:PressurePlots}
\end{figure}

\begin{table}[ht]
\centering
\caption{Nominal and optimized parameter values and the relative change to the nominal estimates after optimization. The wall parameter $\beta$ was the same for all three models in each mouse.}\label{Tab:Optimized} 
 {\footnotesize 
 \setlength\tabcolsep{3.5pt}
 \renewcommand{\arraystretch}{1.2}
 \begin{tabular}{c c c c c c c c}
   \hline\noalign{\smallskip}
  \multicolumn{5}{c}{\bf \hspace{2.2cm}{Control}} & \multicolumn{2}{c}{\bf Hypoxic}\\ [2pt]
  \hline
   Model & Parameter & Nominal & Optimized & Relative & Nominal & Optimized & Relative \\
    &    &    Estimate & Value & Change (\%)&Estimate & Value & Change (\%)\\
   \hline
	- & $\beta$ & $26.0$ & $42.9$ & 65 & $150.9$ & $136.6$ & -10\\
	\hline
	SV & $r_1$  & $1$ & $8.84\times 10^{-1}$ & -12 & 1 & $8.86 \times 10^{-1}$ & -12\\
	SV & $r_2$  & $1$& $1.02$ & 2 & 1 & 1.02 & 2\\
	SV & $c_1$  & $1$ & $1.44$ & 44 & 1 & 1.21 & 21\\
	SV & $R_1/R_T$ & 0.2 & $0.18$ & $-11$ & 0.2 & 0.18 & $ -11$ \\
	SV & $R^{WK}_T$ & $ 78.4$ & $77.6$ & $-1$ & $147 $  & $146$ & $-0.7$ \\
	SV & $C^{WK}_P$ & $ 1.7 \times 10^{-3}$ & $2.5 \times 10^{-3} $ & $44$ & $5.9 \times 10^{-4} $  & $7.1 \times 10^{-4}$ & $21$ \\
	\hline	

	SB & $r_1$  & $1$ & $8.00 \times 10^{-1}$ & -20 & 1 & $8.21 \times 10^{-1}$ & -18\\
	SB & $r_2$  & $1$ & $9.95 \times 10^{-1}$ & -0.5 & 1 & 1.00 & 0.3\\
	SB & $c_1$ & $1$ & $1.34$  & 34 & 1 & 1.14 & 14\\
	SB & $R_1/R_T$ & 0.2 & $0.17$ & $-17$ & 0.2 & 0.17& $ -15$ \\
	SB & $R^{WK}_T$ & $ 78.4$ & $74.9$ & $-4$ & $147$  & $142$ & $-3$ \\
	SB & $C^{WK}_P$ & $ 1.7 \times 10^{-3}$ & $2.3 \times 10^{-3} $ & $34$ & $5.9 \times 10^{-4} $  & $6.7 \times 10^{-4}$ & $14$ \\
	\hline

	FN & $r_1$  & $1$ & $1.34 \times 10^{-1}$ & 87 & 1 & $5.40 \times 10^{-1}$ & 46\\
	FN & $r_2$  & $1$ & $8.82 \times 10^{-1}$  & 12 & 1 & $9.70 \times 10^{-1}$ & 3\\
	FN & $c_1$  & $1$ & $1.62$  & 62 & 1 & 1.07 & 7\\
	FN & $R_1/R_T$ & 0.2 & $0.04$ & $-82$ & 0.2 & 0.12 & $ -39$ \\
	FN & $R^{WK}_T$ & $ 78.4$ & $57.4$ & $-27$ & $147$  & $130$ & $-12$ \\
	FN & $C^{WK}_P$ & $ 1.7 \times 10^{-3}$ & $2.8 \times 10^{-3} $ & $62$ & $5.9 \times 10^{-4} $  & $6.3 \times 10^{-4}$ & $7$ \\
	 \noalign{\smallskip}\hline\\[-6pt]
  \end{tabular}} \\
   {\footnotesize Units: $\beta$ (mmHg), $R^{WK}_T$ (mmHg s/ml), $C^{WK}_P$ (ml/mmHg), and $r_1,r_2,c_1,R_1/R_T$ (dimensionless)}
\end{table}

\subsection{Uncertainty Quantification}

DRAM simulations were initialized using the estimated values from the SQP optimization. The stiffness parameter $\beta$ was fixed, while the scaling parameters $\theta = \{r_1, r_2, c_1\}$ parameters were allowed to vary. Each simulation used a 2,000 iteration burn in period to initialize a 10,000 iteration chain. The bounds for the prior distributions were set to $\pm 50\%$ of the optimized values.  Plots of the densities for all three models are shown in Fig. \ref{fig:Densities}, with initial estimates (the optimized values form the SQP optimization ($\hat{\theta}$)) marked with asterisks on the density curve. 

The maximum density parameter values were used to predict pulmonary arterial pressure (Fig.~\ref{fig:PressurePlots}).  The variance of the parameters, shown qualitatively by the width of the posterior densities, agree with the local sensitivity results. The width of $c_1$ is largest, indicating 
less impact on the model predictions within the parameter space it is sampled from. $r_1$ and $r_2$ are more sensitive, and hence have narrower distributions. The variance estimates for each of the parameters are given in Table~\ref{Tab:Posterior}.

\begin{figure}[ht]
\centering
\includegraphics[width=0.75\textwidth,angle=270]{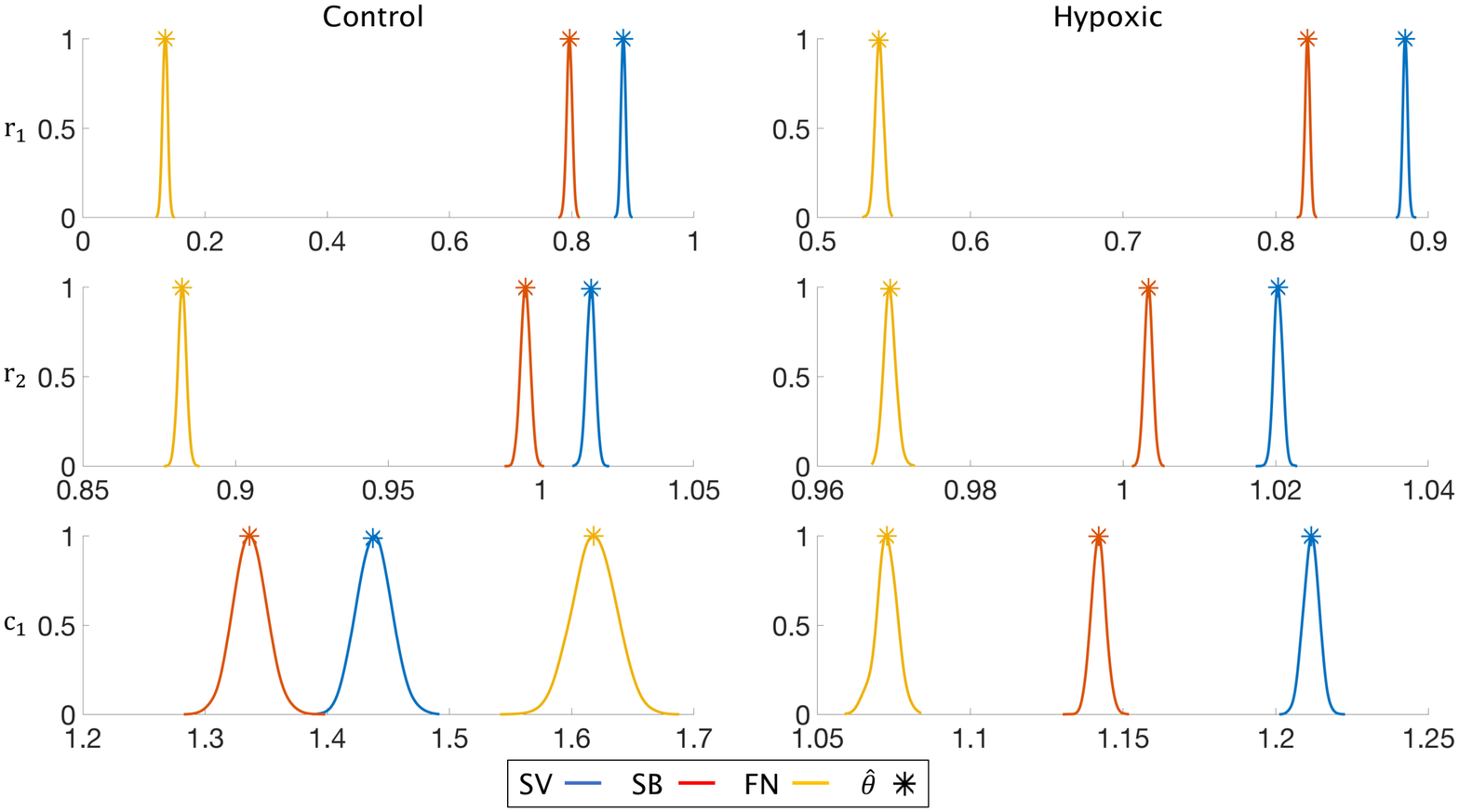}
\vspace{-8mm}
\caption{Posterior densities obtained from 10,000 chain iterations of DRAM of all three models for the control and hypoxic mice after a 2,000 iteration burn in period. The parameter values obtained from the SQP optimization are plotted in asterisks on the density curves.}\label{fig:Densities}
\end{figure}

\begin{table}[ht]
\centering
\caption{Posterior parameter density variance values for the SV, SB, and FN models for both the control and hypoxic mice.}\label{Tab:Posterior} 
 {\footnotesize 
 \setlength\tabcolsep{3.5pt}
 \renewcommand{\arraystretch}{1.2}
 \begin{tabular}{c c c c}
   \hline\noalign{\smallskip}
  \multicolumn{3}{c}{\bf \hspace{2.2cm}{Control}} & \multicolumn{1}{c}{\bf Hypoxic}\\ [2pt]
  \hline\\
   Model & Parameter & Posterior $\sigma_\theta^2$ & Posterior  $\sigma_\theta^2$ \\ \hline
	SV & $r_1$  & $7.93 \times 10^{-5}$ & $1.56 \times 10^{-5}$\\
	SV & $r_2$  & $1.25 \times 10^{-5}$ & $2.54 \times 10^{-6}$\\
	SV & $c_1$  & $9.20 \times 10^{-4}$ & $3.98 \times 10^{-5}$\\
	\hline
	SB & $r_1$  & $1.06 \times 10^{-4}$ & $1.53 \times 10^{-5}$\\
	SB & $r_2$  & $1.45 \times 10^{-5}$ & $1.57 \times 10^{-6}$\\
	SB & $c_1$ & $1.15 \times 10^{-3}$ & $4.05 \times 10^{-5}$\\
	\hline
	FN & $r_1$  & $2.76 \times 10^{-4}$ & $3.04 \times 10^{-4}$\\
	FN & $r_2$  & $9.93 \times 10^{-6}$ & $1.04 \times 10^{-5}$\\
	FN & $c_1$  & $1.85 \times 10^{-4}$ & $1.16 \times 10^{-4}$\\
	\hline
  \end{tabular}}
\end{table}

Confidence intervals for the parameters were calculated around the optimized values $\hat{\theta}$ obtained from the SQP algorithms. The parameter confidence intervals are given in Table \ref{Tab:CI}. The intervals for the control model parameters were larger than that found for the hypoxic parameters, which is due to the larger residual obtained from the control models.

Frequentist prediction intervals were calculated using the optimized values and the optimized parameter sensitivity matrix. The posterior densities from DRAM were used to construct Bayesian credible and prediction intervals for the model response. 1,000 samples from the parameter densities were take  to create the intervals. The confidence, prediction, and credible intervals are shown in Fig. \ref{fig:PressurePlots}. The control mouse intervals are larger than the hypoxic, indicating a greater amount of uncertainty in the control model.
\begin{table}[ht]
\centering
\caption{Frequentist confidence intervals for the optimized Windkessel parameters $\hat{\theta}_{WK} = \{r_1, r_2, c_1\}$ for the control and hypoxic models.}\label{Tab:CI} 
 {\footnotesize 
 \setlength\tabcolsep{3.5pt}
 \renewcommand{\arraystretch}{1.2}
 \begin{tabular}{c c c c c c }
   \hline\noalign{\smallskip}
  \multicolumn{4}{c}{\bf \hspace{1.5cm}{Control}} & \multicolumn{2}{c}{\bf Hypoxic}\\ [2pt]
  \hline\noalign{\smallskip}
   Model & Parameter & Optimized & Confidence & Optimized & Confidence \\
         &   &  Value & Interval & Value & Interval\\
   \noalign{\smallskip}\hline\noalign{\smallskip}
	SV & $r_1$ & $8.84\times 10^{-1}$ & $[8.76 ,8.93]\times10^{-1}$ & $8.85\times 10^{-1}$ & $[8.81, 8.88] \times 10^{-1}$ \\
	SV & $r_2$ & $1.02$ &$[1.01 ,1.02]$ & $1.02$ & $[1.02, 1.02]$ \\
	SV & $c_1$ & $1.44$ &$[1.42, 1.46]$ & $1.21$ & $[1.21, 1.22]$\\
	\hline
	SB & $r_1$ & $7.96\times 10^{-1}$ & $[7.85, 8.08]\times10^{-1}$  & $8.21\times 10^{-1}$ &$[8.17, 8.24]\times10^{-1}$ \\
	SB & $r_2$ & $9.95\times 10^{-1}$ &  $[9.92, 9.98]\times10^{-1}$& $1.00$ &$[1.00, 1.00]$\\
	SB & $c_1$ & $1.34$ &$[1.32, 1.36]$  & $1.14$ &$[1.14, 1.15]$ \\
	\hline
	FN & $r_1$ & $1.35\times 10^{-1}$ & $[0.75, 1.95]\times10^{-1}$ & $5.40\times 10^{-1}$ & $[5.31, 5.49] \times10^{-1}$ \\
	FN & $r_2$ & $8.82\times 10^{-1}$ & $[8.79, 8.86]\times10^{-1}$ & $9.70\times 10^{-1}$ & $[9.68, 9.71]\times 10^{-1}$ \\
	FN & $c_1$ & $1.62$ &$[1.60, 1.64]$  & $1.07$ & $[1.07, 1.08]$ \\
    \noalign{\smallskip}\hline\\[-6pt]    
  \end{tabular}}
\end{table}

\begin{figure}[ht]
\centering
\includegraphics[width=0.75\textwidth,angle=270]{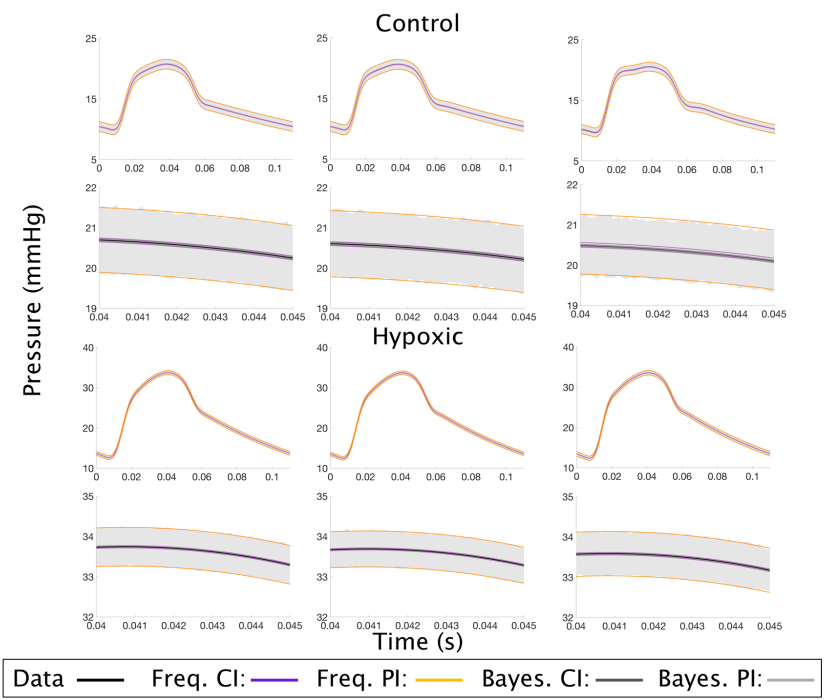}
\caption{Frequentist confidence and prediction intervals for the model response plotted against bayesian credible and prediction intervals for the model response obtained from DRAM simulations.}\label{fig:Intervals}
\end{figure}

\subsection{Network Predictions}

The pressure and flow data are only measured in the main pulmonary artery, which limits the ability to compare model predictions to data in the distal vasculature. However, the model can still predict the pressure and flow in each vessel in the pulmonary tree. Figure \ref{fig:NetworkPredictions} shows model predictions for seven of the vessels in the FN model for both the control and hypoxic mice. 

Model predictions in close proximity to the main pulmonary artery retain the systolic pressure and pulse pressure, in contrast  the distal vessel predictions show a decrease in these values. The flow predictions decrease in magnitude downstream as well. The pulse pressure, the difference between the maximum and minimum pressure, decreased by $68\%$ in the control vs. $28\%$ in the hypoxic mouse, indicating that the pulse pressure in the hypoxic mouse does not dissipate as much as in the control mouse. 

\begin{figure}[ht]
\centering
\includegraphics[width=0.75\textwidth,angle = 270]{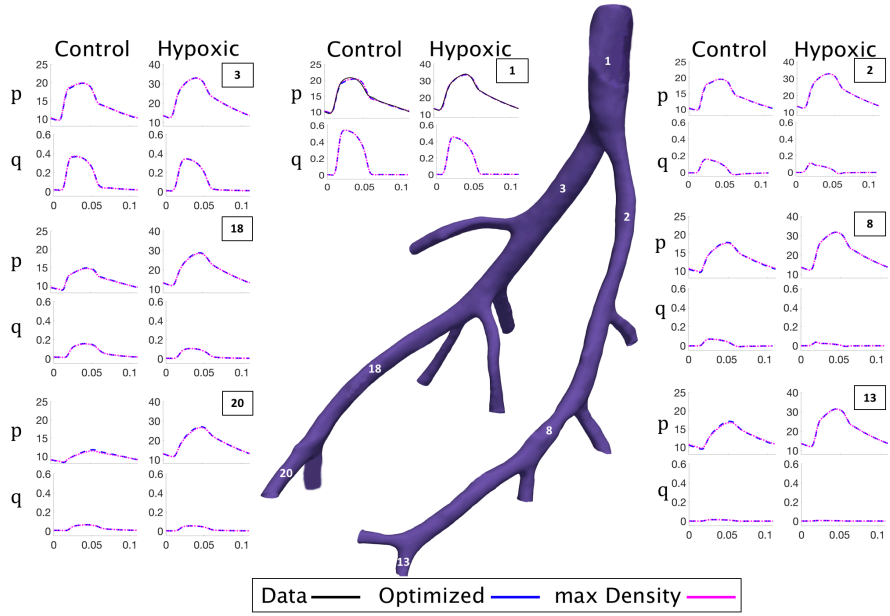}
\caption{Model predictions of pressure and flow in the downstream vasculature using the optimized and maximum density parameter values. Predictions in the left column for each location identifies the control mouse while the right column shows the prediction in the hypoxic mousel.}\label{fig:NetworkPredictions}
\end{figure}

\section{Discussion} \label{sec:3}

Using flow and pressure measurements from the main pulmonary artery of a control and hypoxic  mouse, this study investigated how model parameters change with network size and physiological conditions. We used local and global sensitivity analysis to determine parameter importance and interaction followed by correlation and sampling (DRAM) analysis  to  identify parameter correlations. Subsequently, we used optimization and uncertainty quantification to determine how well the model fits data. Overall, our results show that the peripheral vascular resistance $R_2$ is the most sensitive parameter,  that the large vessel stiffness $\beta$ increases in sensitivity with network size, while the peripheral vascular compliance $C_p$ and the resistance $R_1$ decrease in sensitivity with network size.

The pulmonary circulation consists of an expansive network of blood vessels, which branch in rapid succession from the main pulmonary artery to the capillaries encapsulating the alveoli. It is known that the pulmonary vasculature is modulated by disease, both structurally by changing the network morphometry and materially by changing vessel stiffness. Current classification of pulmonary hypertension and its progression is based on assessing if the pathophysiology is located in the proximal (the large vessels) or peripheral vasculature. To build tools to distinguish the disease sub-classifications and progression, we analyzed how parameter estimates vary with network size. Giving more insight into how to develop a multi scale model that can distinguish the two network components. Given the change in vessel size is gradual, it is not obvious how to distinguish ``large" and ``small" peripheral vessels. Two previous studies have addressed this question analyzing systemic arterial dynamics \cite{Epstein2015,Melis2017}, but to our knowledge  no previous studies have analyzed the pulmonary circulation.  

To study how the sensitivity of parameters change with network size and disease we contructed three representative networks (SV, SB, FN) with different 1D to 0D ratios. Here the 1D model represents the ``large" vessels while the 0D model represents the ``small" peripheral vessels. Results showed that the vessel stiffness $\beta$ becomes more sensitive/important as vessels are added to the network, while the scaling factors $r_1$ and $c_1$ become less sensitive, i.e. it becomes easier to infer $\beta$ and harder to infer $r_1$ and $c_1$. The $r_1$ scales the proximal resistance $R_1$, representing the characteristic peripheral impedance, and $c_1$ scales the peripheral compliance $C_p$. These results indicate that if the number of proximal vessels are large enough it may be possible to apply a simpler boundary condition at the outlets, e.g. a two element Windkessel or a pure resistance model. 

Both the  structured correlation analysis (local) and DRAM (global) showed that for the hypoxic mouse $\beta$ and $r_1$ are highly correlated ($|c_{ij}|>0.95$) and somewhat correlated for the control mouse ($|c_{ij}| > 0.75$). Initial optimization results allowing $\beta$ to vary with network size reflected the correlation, i.e. for some networks $\beta$ was large, while $r_1$ was small, but for others the result was opposite. Yet for a given hemodynamic condition, the vascular model should have constant material properties irrespective of the model complexity. To improve the nominal estimate for $\beta$ we set up optimization scheme over all three networks for each mouse, where we estimated a common value of $\beta$ along with estimates of $r_1, r_2, c_1$ for each network size. 

For the sampling based method (DRAM) we kept $\beta$ constant at optimized value. Ideally, DRAM should have been run for all networks, but this simulation was too computational intensive. Results of the combined approach allowed us to obtain better fits than by (a) keeping $\beta$ fixed at its nominal value and (b) estimating all parameters.  As expected, our results show that the hypoxic mouse has a lower peripheral compliance than the control mouase, and that peripheral compliance and large vessel stiffness do not change significantly with network size. This indicates that the assumption of constant stiffness in the largest arteries for modeling purposes is reasonable. 

The DRAM results and uncertainty quantification supported the results from the least squares parameter estimation. The overlap between the maximum density posterior estimates and the optimization indicates that no other local minima exist within the parameter bounds sampled. Likewise, the prediction intervals indicate that both the control and hypoxic models are accurate, as the $95\%$ prediction intervals shows only a $\pm1$ mmHg band around the data. Further analysis of the parameter distributions showed that the control mouse has wider bands around pressure prediction, indicating a more complex physiology. Only a few studies have been carried out to determine the effects of \textit{aleatoric} uncertainty, which includes uncertainty in measured geometry from imaging modalities \cite{Sankaran2015,Melis2017}, which should be investigated further.

Results discussed here illustrate how to incorporate sensitivity analysis, subset selection, optimization, and uncertainty quantification to study dynamics, yet results clearly depend on the model analyzed and data available for model validation. Major limitations of the model analyzed here include the assumption that the same constitutive law can be applied to analyze data from the control and hypoxic mice. Our results show that the assumption is valid for the hypoxic model where wall remodeling has likely resulted in stiffer walls, but that the model could be improved for the control mouse. This agrees with previous findings that arterial vessel deformation is nonlinear and viscoelastic~\cite{Steele11,Valdez11,Lee2016}. 

Another limitation is that the assumption of iid errors is  violated due to the numerous parameter interactions in the  model. This assumption was made for simplicity in the optimization and MCMC routines, a plot of the residuals in Fig. \ref{fig:Residuals} shows that the residuals are in fact not independent. To tackle this violation, one could instead employ the log-likelihood $\log\left(\cal{{L}}\right)$, defined as
\BE \label{eq:LogLikelihood}
\log\left(\cal{{L}}\right) = -\frac{1}{2}\log\left(\det\left(2\pi\Sigma\right)\right) - \frac{1}{2} r^T \Sigma^{-1} r,z1
\EE
where $\Sigma$ is the covariance matrix between parameters. 
\begin{figure}
\centering
\includegraphics[width=0.75\textwidth]{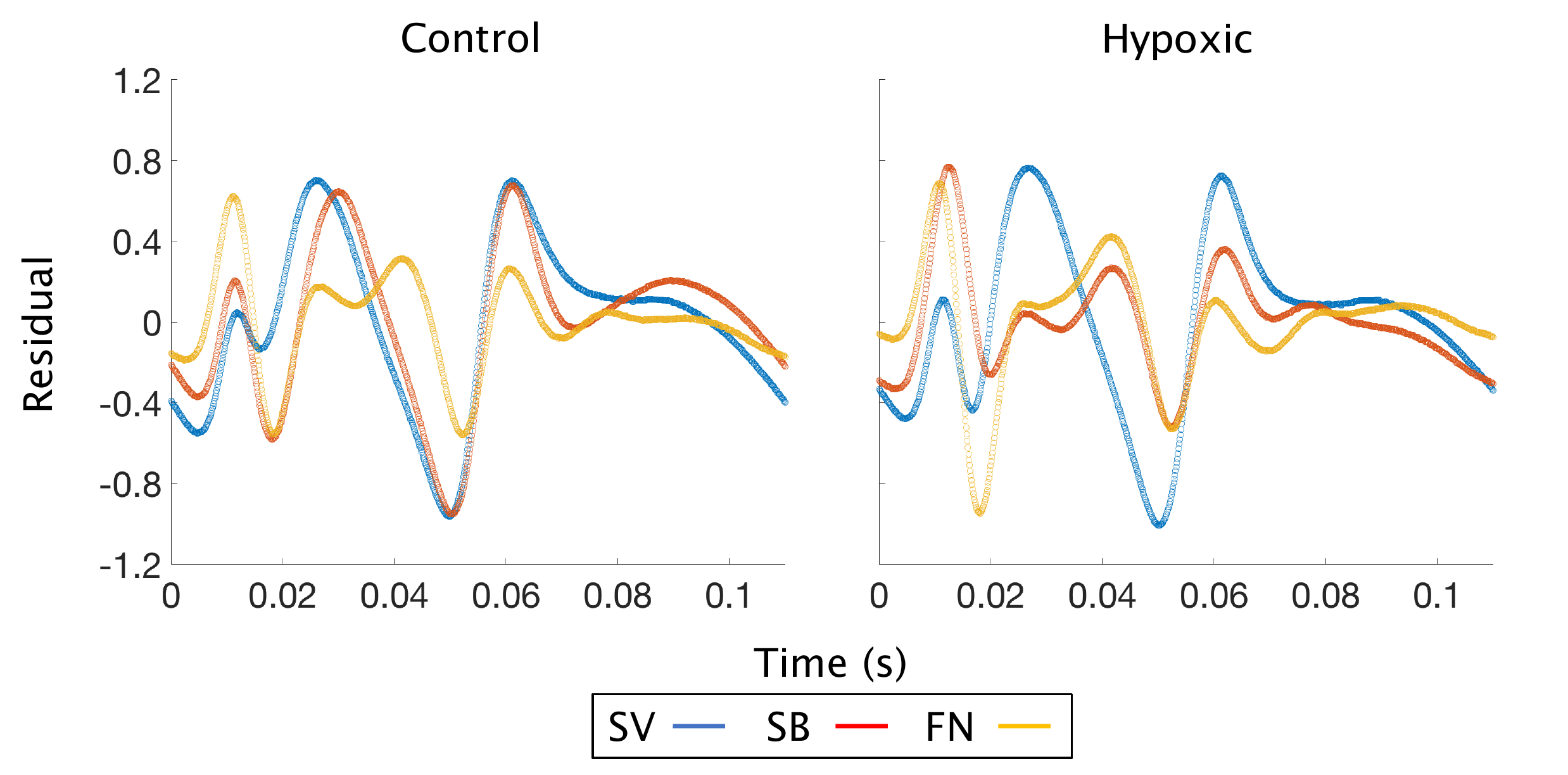}
\caption{Residuals from the model predictions with the optimized parameters. The residual curves indicate that errors are not independent, violating the simplifying assumptions often made about physical models.}\label{fig:Residuals}
\end{figure}

Finally, we only investigated impact on vessel stiffness and outflow boundary condition parameters fixing dimensions extracted from imaging studies. Clearly, variation in unstressed vessel radius impacts model predictions, as does the assumption of constant viscosity, which for the large network could likely vary due to Fahraeus Lindqvist effect \cite{Lighthill75}. Moreover, results presented here are limited by the fact that data was only available in the main pulmonary artery. The methods proposed here can easily be expanded to study any of these factors.

\section{Conclusion} \label{sec:4}

The goal of this study was to analyze a coupled 1D-0D model of pulse wave propagation in the pulmonary circulation. We analyzed parameter sensitivity and correlation, estimated identifiable model parameters, determined uncertainty intervals, and studied the parameter effect on changing network complexity. To fit the model to data we estimated vessel stiffness and global scaling parameters adjusting Windkessel parameters for two mice (control and hypoxic) and three networks of varying complexity.   Results showed that the hypoxic mouse has stiffer proximal $\beta$ and peripheral $R_2$ vessels. Moreover we showed that sensitivity of the proximal vessel stiffness increase with vessel size relative to $R_1$ (the proximal resistance in the Windkessel model) and $C_p$ (the peripheral compliance).  The observation, that the parameters effect on the model output (pulmonary arterial pressure) vary with network complexity is essential to account for when developing models that delineate proximal vs. peripheral vessels to study disease classification and progression.

\section{Acknowledgements}
We thank Prof. Naomi Chesler at the University of Wisconsin, Madison for sharing the measured waveforms and micro-CT images. We thank Prof. Dirk Husmeier and L. Mihaela Paun at the University of Glasgow, Scotland for discussing how to set up Bayesian analysis. We thank Prof. Ralph Smith at NC State University for discussing how to approach global sensitivity analysis. We thank Tina Ghashghaei, Apex High School for help with artwork in Fig. 1.

\section{Funding}
This study was supported by the National Science Foundation (NSF) awards NSF-DMS \# 1615820 and NSF-DMS\# 1246991.

\section{Animal Studies}
This study analyzes existing hemodynamic and micro computed tomography (micro-CT) imaging data from a control and a hypoxia induced hypertensive mouse extracted from a group of 7 control and 5 hypoxic animals, respectively. Detailed experimental protocols describing this data can be found in \cite{Tabima2012,Vanderpool2011}. All experimental procedures are approved by the University of Wisconsin Institutional Animal Care and Use Committee. Data stripped of identifiers were made available by Prof. Naomi Chesler, University of Wisconsin, Madison.
The authors were not involved in performing any of the animal studies.

\section{Conflict of Interest}
Mitchel J. Colebank, M. Umar Qureshi, and Mette S. Olufsen all declare that they have no conflicts of interest.

\section{Ethical Approval}
This article does not contain any studies with human participants performed by any of the authors.

% Non-BibTeX users please use


\newpage
\begin{thebibliography}{}

\bibitem{Alalstruey08}
Alastruey, J., Parker, K.H., Peiro, J., Sherwin, S.J. 
Lumped parameter outflow models for 1D blood flow simulations: effect on pulse waves and parameter estimation. 
Commun Comput Phys 4:317--336 (2008)

\bibitem{Alexanderian2017}
Alexanderian, A., Gremaud, P. A., Smith, R.C. 
Variance-based sensitivity analysis for time-dependent processes.
Stat 	arXiv 1711.08030 (2017)

\bibitem{VMTK}
Antiga, L., Piccinelli, M., Botti, L., Ene-Iordache, B., Remuzzi, A., Steinman, D.A.
An image-based modeling framework for patient-specific computational hemodynamics.
Med Biol Eng Comput 46:1097-1112 (2008)
%(http://www.vmtk.org). 

\bibitem{Arnold17}
Arnold, A., Battista, C., Bia, D., Z—calo German, Y., Armentano, R.L., Tran, H.T., Olufsen, M.S.
Uncertainty quantification in a patient-specific 1D arterial network model: EnKF-based Inflow estimator.
J Verif Valid Uncert 2:011002 - 14 pages (2017)

\bibitem{Banks2013}
Banks, H.T., Cintron-Arias A. Kappel, F.
Parameter selection methods in inverse problem formulation. In
Mathematical modeling and validation in physiology: applications to the cardiovascular and respiratory systems.
Springer, Berlin Heidelberg, Germany (2013)

\bibitem{Banks2015}
Banks, H.T., Bekele-Maxwell, K., Bociu, L., Noorman, M., Tillman, K.
The complex-step method for sensitivity analysis of non-smooth problems arising in biology.
Eurasian J Math Comput Appl 3:1--41 (2015)
 
\bibitem{Battista16}
Battista, C., Bia, D., Zocalo German, Y., Armentano, R.L.,  Haider, M.A., Olufsen M.S.
Wave propagation in a 1D fluid dynamics model using pressure-area measurements from ovine arteries. 
J Mech Med Biol 16:1650007 - 26 pages (2016)
     
\bibitem{Brady18}
Brady, R., Frank-Ito, D.O., Tran, H.T., Janum, S., Muller, K., Brix, S., Ottesen, J.T., Mehlsen, J. Olufsen, M.S. 
Personalized mathematical model of endotoxin-induced inflammatory responses in young men and associated changes in heart rate variability. 
Math Model Nat Phenom, in Press (2018)

\bibitem{Brault16}
Brault, A., Dumas, L., Lucor, D. 
Uncertainty quantification of inflow boundary condition and proximal arterial stiffness coupled effect on pulse wave propagation in a vascular network.
Int J Numer Method Biomed Eng  33:e2859 (2016)

\bibitem{Campolongo2007}
Campolongo, F., Cariboni, J., Saltelli, A.
An effective screening design for sensitivity analysis of large models
Environ Model Softw 22:1509--1518 (2007)

\bibitem{Eck16}
Eck, V. G., Donders, W. P., Sturdy, J. Feinberg, J. Delhaas, T., Hellevik, L.R., Huberts, W.
Advances in 0D and 1D models for circulation: A guide to uncertainty quantification and sensitivity analysis for cardiovascular applications. 
Int J Numer Meth Biomed Eng  32:e02755 (2016)

\bibitem{Ellwein2008}
Ellwein, L.M., Tran, H.T., Zapata, C., Novak, V., Olufsen, M.S. 
Sensitivity analysis and model assessment: Mathematical models for arterial blood flow and blood pressure. 
Cardiovasc Eng 8:94--108 (2008)

\bibitem{Ellewin2015}
Ellwein, L.M., Marks, D.S., Migrino, R.Q., Foley, W.D., Sherman, S., LaDisa, J.F.
Image-based quantification of 3D morphology for bifurcations in the left coronary artery: application to stent design.
Catheter Cardiovasc Interv 87:1244--1255  (2016)

\bibitem{Epstein2015}
Epstein,S., Willemet,M, Chowienczyk, P.J., Alastruey, J. 
Reducing the number of parameters in 1D arterial blood flow modeling: less is more for patient-specific simulations. 
Am J Physiol 309:H222-H234 (2015)

\bibitem{Feldkamp1984}
Feldkamp, L.A., Davis, L.C., Kress, J.W.
Practical cone-beam algorithm.
J Opt Soc Am A 1:612--619 (1984)
  
\bibitem{Griewank1989}
Griewank, A. 
On automatic differentiation.
Math program: recent developments and applications  6: 83--107 (1989)

\bibitem{Grinberg08}
Grinberg, L, Karniadakis, G.E. 
Outflow boundary conditions for arterial networks with multiple outlets.
Ann Biomed Eng 36:1496--1514 (2008)

\bibitem{Gul16}
Gul, R., SchŸtte, C., Bernhard, S. 
Mathematical modeling and sensitivity analysis of arterial anastomosis in the arm. 
Appl Math Model 40:7724--7738  (2016)
  
\bibitem{Haario2006}
Haario, H., Laine, M., Mira, A., \& Saksman, E.  
DRAM: Efficient adaptive MCMC
Stat Comput 16:339--354 (2006)
%https://doi.org/10.1007/s11222-006-9438-0
  
\bibitem{Lott1985}
Iott, J., Haftka, R.T., Adelman, H.M.
Selecting step sizes in sensitivity analysis by finite differences.
NASA Tech Memo NASA-TM-86382, L-15938, NAS 1.15:86382 (1985)

\bibitem{Karau2001}
Karau, K., Johnson, R., Molthen, R., Dhyani, A., Haworth, S., Hanger, C., Roerig, D., Dawson, C.
Microfocal X-ray CT imaging and pulmonary arterial distensibility in excised rat lungs.
Am J Physiol 281:H1447--H1457 (2011)

\bibitem{Krenz2003}
Krenz, G., Dawson, C.
Flow and pressure distributions in vascular networks consisting of distensible vessels.
Am J Physiol 284:H2192--H2203 (2003).

\bibitem{Laine2007}
Laine, M. 
MCMC Toolbox for Matlab. 
http://helios.fmi.fi/ lainema/dram/ (2007)

\bibitem{Lee2016}
Lee, P., Carlson, B.E., Chesler, N., Olufsen, M.S., Qureshi, M.U., Smith, N.P., Sochi, T.,  Beard, D.A. Heterogeneous mechanics of the mouse pulmonary arterial network
Biomech Model Mechanobiol 15:1245--1261 (2016)

\bibitem{vandeVosse11}
Leguya, C.A.D, Bosboom, E.M.H., Belloum, A.S.Z., Hoeks, A.P.G. van de Vosse, F.N.
Global sensitivity analysis of a wave propagation model for arm arteries. 
Med Eng Phy 33:1008--1016 (2011)

\bibitem{Lighthill75}
Lighthill, Sir J. 
Mathematical biofluid dynamics.
SIAM, Philadelphia, PA (1975)

\bibitem{Marquis18}
Marquis, A.D. Arnold, A., Dean, C., Carlson, B.E., Olufsen, M.S. 
Practical identifiability and uncertainty quantification of a pulsatile cardiovascular model. 
Q-bio arXiv 1710.07989 (2018)

\bibitem{McDonald:UK} 
McDonald, D.A., Attinger, E.O.
The characteristics of arterial pulse wave propagation in the dog. 
Inform Exch Gp. No 3, Sci Mem 7 (1965)

\bibitem{Melis2017}
Melis, A., Clayton, R.H., Marzo, A.
Bayesian sensitivity analysis of a 1D vascular model with Gaussian process emulators
Int J Numer Method Biomed Eng  33:e2882 (2017)

\bibitem{Miao11}
Miao, H., Xia, X., Perelson, A.S., Wu, H. 
On identifiability of nonlinear ODE models and applications in viral dynamics. 
SIAM Rev 53:3--39 (2011) 

\bibitem{Mirams16}
Mirams, G.R., Pathmanathan, P., Gray, R.A., Challenor, P., Clayton, R.H. 
Uncertainty and variability in computational and mathematical models of cardiac physiology.
J Physiol 594.23: 6833D6847 (2016)

\bibitem{Morris1991}
Morris, M.D.
Factorial plans for preliminary computational experiments
Technometrics  33:161--174 (1991)

\bibitem{Mynard15}
Mynard, J.P., Smolich, J.J. 
One-dimensional haemodynamic modeling and wave dynamics in the entire adult circulation. 
Ann Biomed Eng 43:1443-1460 (2015)
%DOI: 10.1007/s10439-015-1313-8

\bibitem{Olsen2015}
Olsen, C., Tran, H., Ottesen, J. T., Mehlsen, J., Olufsen, M. 
Challenges in practical computation of global sensitivities with application to a baroreceptor reflex model.
NCSU CRSC reports CRSC-TR13-15 (2013)

\bibitem{Olufsen2000}
Olufsen, M.S., Peskin, C.S., Kim, W.Y., Pedersen, E.M., Nadim, A., Larsen, J.
Numerical simulation and experimental validation of blood flow in arteries with structured-tree outflow conditions.
Ann Biomed Eng 28:1281--1299 (2000)

\bibitem{Olufsen2012}
Olufsen, M.S., Hill, N.A., Vaughan, G.D., Sainsbury, C., Johnson, M.
Rarefaction and blood pressure in systemic and pulmonary arteries.
J Fluid Mech 705:280--305 (2012)

\bibitem{Olufsen13}
Olufsen, M.S., Ottesen, J.T. 
A practical approach to parameter estimation applied to model predicting heart rate regulation. 
J Math Biol 67:39--68 (2013)

\bibitem{Paun2018}
Paun, L.M., Qureshi, M.U., Colebank, M., Hill, N.A., Olufsen, M.S., Haider, M.A., Husmeier, D.
MCMC methods for inference in a mathematical model of pulmonary circulation.
Statistica Neerlandica doi.org/10.1111/stan.12132 (2018)
 
\bibitem{Pope2009}
Pope, S., Ellwein, L., Zapata, C., Novak, V., Kelley, C., Olufsen, M. 
Estimation and identification of parameters in a lumped cerebrovascular model. 
Math Biosci Eng 6:93--115 (2009)

\bibitem{Quateroni16}
Quarteroni, A.,Veneziani, A., Vergara, C. 
Geometric multiscale modeling of the cardiovascular system, between theory and practice. 
Comput Methods Appl Mech Eng 302:193--252  (2016) 

\bibitem{Qureshi2014}
Qureshi, M.U., Vaughan, G.D., Sainsbury, C., Johnson, M., Peskin, C.S., Olufsen, M.S., Hill, N.A.
Numerical simulation of blood flow and pressure drop in the pulmonary arterial and venous circulation.
Biomech Model Mechanobiol 13:1137--1154 (2014)
 
\bibitem{Qureshi2017}
Qureshi, M.U., Colebank, M.J., Schreier, D.A., Tabima, D.M., Haider, M.A., Chesler, N.C., Olufsen, M.S.
Characteristic impedance: frequency or time domain approach?
Physiol Meas 39:014004  (2018)

\bibitem{Qureshi2018}
Qureshi, M.U., Colebank, M.J., Paun, M., Ellwein, L., Chesler, N., Haider, M.A., Hill, N.A., Husmeier, D., Olufsen, M.S. 
Hemodynamic assessment of pulmonary hypertension in mice, A model based analysis of the disease mechanism. 
Physics arXiv 1712.01699 (2018)

\bibitem{Raghu11}
Raghu, R., Vignon-Clementel, I.E., Figueroa, C.A., Taylor, T.A.  
Comparative study of viscoelastic arterial wall models in nonlinear one-dimensional finite element simulations of blood flow. 
J Biomech Eng, 133:081003 - 11 pages (2011)
%doi:10.1115/1.4004532

\bibitem{Reymond2009}
Reymond, P., Merenda, F., Perren, F., Rufenacht, D., Stergiopulos, N.
Validation of a one-dimensional model of the systemic arterial tree.
Am J Physiol 297: H208--H222 (2009)
  
\bibitem{Riches1973}
Riches, A.C., Sharp, J.G., Thomas, D.B., Smith, S.V.
Blood volume determination in mouse.
J Physiol 228:279--284 (1973)
  
\bibitem{Saltelli2008}
Saltelli, A., Ratto, M., Andres, T., Campolongo, F., Cariboni, J., Gatelli, D., Saisana, M., Tarantola, S. Global sensitivity analysis. The primer.
John Wiley and Sons, Chichester, UK (2008)

\bibitem{Sankaran2015}
Sankaran, S., Grady, L., Taylor, C.A.
Fast computation of hemodynamic sensitivity to lumen segmentation uncertainty
IEEE Trans Med Imaging 34:2562-2571 (2015)

\bibitem{Safaei2016}
Safaei, S., Bradley, C.P., Suresh, V., Mithraratne, K., Muller, A., Ho, H., Ladd, D., Hellevik, L.R., Omholt, S.W., Chase, J.G., Muller, L.O., Watanabe, S.M., Blanco, P.J., de Bono, B., Hunter, P.J. Roadmap for cardiovascular circulation model. 
J Physiol 594:6909--6928 (2016)

\bibitem{Shi11}
Shi, Y., Lawford, P. Hose, R. 
Review of 0D and 1D models of blood flow in the cardiovascular system. 
BioMed Eng Online 10:33--38 (2011)
%http://www.biomedical-engineering-online.com/content/10/1/33

 \bibitem{Steele11}
Steele, B.N., Valdez-Jasso, D., Haider, M.A., Olufsen, M.S. 
Predicting arterial flow and pressure dynamics using a 1D fluid dynamics model with a viscoelastic wall. 
SIAM J Appl Math 71:1123--1143 (2011)

\bibitem{RalphBook}
Smith, R.C.
Uncertainty quantification: theory, implementation, and applications. 
SIAM, Philadelphia, PA  (2014)

\bibitem{Stergiopulos1995}
Stergiopulos, N., Meister, J.J., Westerhof, N.
Evaluation of methods for estimation of total arterial compliance. 
Am J Physiol 268:H1540--1548  (1995)

\bibitem{Sumner2012}
Sumner, T. and Shephard, E. and Bogle, I. D. L.
A methodology for global-sensitivity analysis of time-dependent outputs in systems biology modelling.
J R Soc Interface, 74:2156-2166 (2012)

\bibitem{Tabima2012}
Tabima, D.M., Roldan-Alzate, A., Wang, Z., Hacker, T.A., Molthen, R.C., Chesler, N.C.
Persistent vascular collagen accumulation alters hemodynamic recovery from chronic hypoxia.
J Biomech 45:799--804  (2012)

\bibitem{Valdez11}
Valdez-Jasso, D.,  Bia, D., Z—calo, Y. Armentano, R.L., Haider, M.A., Olufsen, M.S.
Linear and nonlinear viscoelastic modeling of aorta and carotid pressure-area dynamics under in vivo and ex vivo conditions. 
Ann Biomed Eng 39:1438--1456 (2011)

\bibitem{Vanderpool2011}
Vanderpool, R.R., Kim, A.R., Chesler, N.C.
Effects of acute Rho kinase inhibition on chronic hypoxia-induced changes in proximal and distal pulmonary arterial structure and function.
J Appl Physiol 110:188--198 (2011)

\bibitem{Vosse2011} 
van de Vosse, F.N., Stergiopulos, N.
Pulse wave propagation in the arterial tree.
Ann Rev Fluid Mech 43:467--499 (2011)

\bibitem{Wentworth2016}
Wentworth, M.T., Smith, R.C., Banks, H.T. 
Parameter selection and verification techniques based on global sensitivity analysis illustrated for an HIV model. 
SIAM/ASA J Uncert Quant 4:266--297 (2016)

\bibitem{Windberger2003}
Windberger, U., Bartholovitsch, A., Plasenzotti, R., Korak, K.J., Heinze, G.
Whole blood viscosity, plasma viscosity and erythrocyte aggregation in nine mammalian species: reference values and comparison of data.
Exp Physiol 88:431--440 (2003)

\bibitem{Wu18}
Wu, J., Dhingra, R., Gambhir, M., Remais, J.V.  
Sensitivity analysis of infectious disease models: methods, advances and their application. 
J R Soc Interface doi:10.1098/rsif.2012  - 14 pages (2018) 
%http://dx.doi.org/10.1098/rsif.2012.1018

\bibitem{ITKSNAP}
Yushkevich, P.A., Piven, J., Hazlett, H.C., Smith, R.G., Ho, S., Gee, J.C., Gerig, G.
User-guided 3D active contour segmentation of anatomical structures: Significantly improved efficiency and reliability. 
Neuroimage 31:1116--1128 (2006)
%(www.itksnap.org)

\end{thebibliography}
\end{document}